\documentclass[preprint,12pt]{elsarticle}

\usepackage{graphicx} 
\usepackage[utf8]{inputenc}
\usepackage{url}
\usepackage{booktabs}
\usepackage[table,xcdraw]{xcolor}
\usepackage{enumerate}
\usepackage{float}
\usepackage{multirow}
\usepackage{adjustbox}
\usepackage{amsmath, amsthm, amssymb}

\newtheorem{prop}{Proposition}
\newtheorem{obser}{Observation}
\theoremstyle{definition}
\newtheorem{definition}{Definition}
\newtheorem{example}{Example}
\date{February 5, 2017}
\journal{Applied Soft Computing}

\begin{document}
\begin{frontmatter}

\title{A 2-Tuple Fuzzy Linguistic Delphi method to test by consensus content validity of a questionnaire: a case study for b-learning}

\author[label1,label3]{Rosana Montes\corref{cor1}}
\author[label3]{Jeovani Morales}
\author[label3]{Noe Zermeño}
\author[label3]{Jeronimo Duran}
\author[label2,label3]{Francisco Herrera}
\cortext[cor1]{Corresponding author.\\
E-mail addresses: rosana@ugr.es (R. Montes), jeovani@correo.ugr.es (J. Morales), nzermeno@correo.ugr.es (N. Zermeño), jeronimoduran@correo.ugr.es (J. Duran), herrera@decsai.ugr.es (F. Herrera).}

\address[label1]{Department of Software Engineering, University of Granada, Spain}
\address[label2]{Department of Computer Science and Artificial Intelligence, University of Granada, Spain}
\address[label3]{Andalusian Research Institute in Data Science and Computational Intelligence, DaSCI, Spain}

\markboth{R. Montes}{A 2-Tuple Fuzzy Delphi method}


\begin{abstract} 

Classic Delphi and Fuzzy Delphi methods are used to test content validity of a data collection tools such as questionnaires. Fuzzy Delphi takes the opinion issued by judges from a linguistic perspective reducing ambiguity in opinions by using fuzzy numbers. We propose an extension named 2-Tuple Fuzzy Linguistic Delphi method to deal with scenarios in which judges show different expertise degrees by using fuzzy multigranular semantics of the linguistic terms and to obtain intermediate and final results expressed by 2-tuple linguistic values. The key idea of our proposal is to validate the full questionnaire by means of the evaluation of its parts, defining the validity of each item as a Decision Making problem. Taking the opinion of experts, we measure the degree of consensus, the degree of consistency, and the linguistic score of each item, in order to detect those items that affect, positively or negatively, the quality of the instrument. Considering the real need to evaluate a b-learning educational experience with a consensual questionnaire, we present a Decision Making model for questionnaire validation that solve it. Additionally, we contribute to this consensus reaching problem by developing an online tool under GPL v3 license. The software visualizes the collective valuations for each iteration and assists to determine which parts of the questionnaire should be modified to reach a consensual solution.
\end{abstract}



\end{frontmatter}

\section{Introduction}\label{section1}
Quality research must pay attention to the quality of every research process conducted~\cite{berk1990importance}. Data collection procedures are key because the following activities rely heavily on this early step. 
Questionnaires are the most used data collection tools, along with interviews and observation. A desirable property for quality in a questionnaire is the ability to measure the variables for which it was designed, that it is, its validity. \emph{Content validity} is one of three main types of validity evidence. This validity can be tested with the classic Delphi and the Fuzzy Delphi (FD) methods, by taking the consensual opinion of a panel of experts, or judges~\cite{CHANG2000,dalkey1963,Maha76}.

A relatively very popular pedagogical methodology for blending learning (b-learning) is known as \textit{flipped classroom} (FC)~\cite{BERGMANN2012flip}. It is based on \emph{flipping} moments of learning, conceptual acquisition and application of knowledge, allowing students to learn theory outside the classroom, through resources provided by the teacher, mainly videos. Learning happens in different moments: face-to-face as in traditional education \emph{blended} with online activities. Thanks to technological advances that promote interaction between students, the traditional focus of education shifts from individual to collaborative approaches by using technology. Another pedagogical methodology that uses technology in education (mostly mobile) is Mobile Learning or \emph{m-learning} (mL)~\cite{JALDEMARK2017editorial}. Both methodologies are very recent and have attracted by separate the interest of many researchers~\cite{AlEmran18,Karabulut18}. However it is a challenge to apply and to evaluate FC and mL methodologies in combination in a blended learning environment. It is even more because of the lack of standardized questionnaires that address both methodologies in combination. 

We were aware of this difficulty after finding our own need. In the context of undergraduate studies in Computer Engineering at the University of Granada a pilot experience FC \& mL was carried out in a first-year course. We want to evaluate the experience of virtual communication between students-students and teacher-students using the Telegram app via a questionnaire but it should be test for content validity before its application with the students. 

There is a risk of having items or questions that do not target the dimension of interest, or they are badly wording (easy to misread), or they are simply not helpful. It is therefore desirable to test a questionnaire by individually test the items that comprise it. Content validation by expert judgments can be considered as a Multi-Expert Multi-Criteria Linguistic Decision Making (MEMCLDM) problem where people have to choose from many alternatives, and those alternatives are evaluated by a group of experts with regard to some set of criteria by using linguistic variables. 

The application of the Computing with Words (CW) methodology in Decision Making~\cite{mendel2010computing,RUBIN1999,ZADEH1996,zadeh2012computing} has made possible during decades to incorporate linguistic concepts into applied intelligent computer systems~\cite{Montes15,Montes18,morente18}. By nature it is easier for humans to give opinion in natural language than in numerical language. In this sense, the use of linguistic terms rather than numerical assessments is usually more appropriate. For instance, when experts try to evaluate the \emph{usability} of a website, terms such as \emph{good}, \emph{very good} or \emph{bad} are generally used~\cite{ORFANOU2015perceived}; whereas \emph{fast}, \emph{very fast}, or \emph{slow} are used to indicate the \emph{speed} of a car~\cite{chen1992fuzzy}. There is little literature involving questionnaires and Linguistic Decision Making (LDM) problems, though in \cite{CARRASCO2011linguistic} a LDM scenario was set to normalize the results of various questionnaires in the context of different universities, allowing the comparison of the collected data between institutions.

Our proposal is based on obtaining the linguistic opinion of judges in an iterative process for assessing reliance and consensus among the items of the instrument. We call it the 2-Tuple Fuzzy Linguistic Delphi (2TFLD) method which is an extension of the Fuzzy Delphi (FD) with linguistic information represented by the 2-tuple linguistic model. It is used to test content validity of a questionnaire for a experience in b-learning assisted by a software tool which is open, free and online.  

In summary, we intend to achieve the following contributions:
\begin{itemize}


\item Our method can be used to test content validity of a questionnaire, a property that must be satisfied by any data collection tool. 

\item In education, new trends and pedagogical methodologies supported by technology require the design of adapted questionnaires. We describe in Section~\ref{section2} a proposal for a questionnaire designed specifically to concept \emph{satisfaction with} the combined use of FC and ML methodologies in Higher Education and the virtual communication that came about. 

\item To find a consensual solution in a small number of iterations, we compute many linguistic scores for each item as the results of a MEMCLDM model. To provide flexibility in the application of the 2TFLD method, we use Extended Linguistic Hierarchies and expertise degrees associated with dimensions of the questionnaire. Thus we allow multigranular scenarios --more enriched than the typical binary scales-- and unbalance experience of the judges in the research areas associated with the theoretical basis of the questionnaire.

\item We validated the questionnaire for b-learning helped by our software tool in Section~\ref{section5}. After two iterations of 2TFLD method, the consensual questionnaire can be applied to gather the opinion of the students from the course \emph{Fundamentals of Software} at the first semester of the Computer Engineering degree.

\item The 2TFLD online service is a helping tool for the moderator along the steps in the delphi application, but it is also an informative tool for the researcher that, as an expert, has to reach an appropriate degree of consensus with others.

\end{itemize} 

The paper is structured as follows. Section~\ref{section2} reflects on the objectives for questionnaires validation and the design of a questionnaire to be use in b-learning environments. Section~\ref{section3} introduces a 2-Tuple Fuzzy Linguistic method as an extension of both the Fuzzy Delphi and the classical Delphi method. Section~\ref{section4} describes the software tool that supports the moderator in the task of adapting the questionnaire to reach a consensual result in few iterations. Section~\ref{section5} analyzes the results of the the case study that validates a questionnaire for b-learning with an expert panel of nine judges. Finally the paper is concluded in Section~\ref{section6}.

\section{The design of a quesionnaire for b-learning}\label{section2}
This section details the characteristics of an instrument for data collection and the criteria that must be validated in Section~\ref{section21}. We have conducted a piloting experience that we need to measure in terms of satisfaction and virtual communication.
Firstly we describe the methodology used in Section~\ref{section22}. Secondly we explain how we applied FC and mL methodologies in the course \emph{Fundamentals of Software} in Section~\ref{section23}. Finally, the questionnaire is described in Section~\ref{section24}.

\subsection{Properties of a valid questionnaire}\label{section21}

The use of data collection techniques, such as interviewing and questionnaires, are very common in scientific research. Questionnaires are the most widely applied. In fact, there are standard questionnaires for standardized procedures, but innovation is forcing researchers to redesign them so that they can solve the new challenges and can be adapted to new trends.

The design of a questionnaire is not an easy task since it requires a methodical process of design and validation. To define a questionnaire, you have to choose the structure and the questions or item composition. Each item has a description (the text that the user reads), a type (open or closed question) and in the case of closed questions, a scale (yes/no answer or Likert style). More importantly, items are grouped by dimensions trying to target a theoretical construct. Dimensions are intended to be clearly and concisely collected and compared, according to standard of quality criteria. After all of these decisions, a valid data collection tool must meet the following properties: 

\begin{enumerate}[i]
	\item \textit{Reliability}. It refers to the trustworthiness of the data collected and it is related to internal consistency and accuracy~\cite{LISSITZ2007suggested}. Usually it is calculated using the Cronbach's alpha~\cite{CRONBACH1951}, which is a measure based on the average interrelationship under the same construct. The Cronbach's alpha is a value between $0$ and $1$, being desirable to be above $0.70$~\cite{PONTEROTTO2007}. In addition, to test the ability to discriminate items, the Pearson's correlation index measures the extent to which it contributes to the internal consistency of the questionnaire. Items with homogeneous Pearson's indexes (that it is, below to $0.2$) can be eliminated in a trim procedure, which is a frequent operation for tuning a questionnaire.
	\item \textit{Validity}. It is the capacity of an instrument to measure the variable for which it is designed. This property is subdivided in three sub-dimensions:
	 \begin{itemize}
	 	\item \textit{Criterion validity}. Refers to the degree of effectiveness by which the variable of interest is predicted with the test scores. Therefore, the operationalization of the concept is based on the validity coefficient, which is the correlation between test and criterion. The higher the correlation, the greater the predictive capacity of the test. There are different designs to determine the correlation of the criterion with the test: (1) concurrent or simultaneous validity, (2) forecast or predictive validity, (3) retrospective validity \cite{evers2013assessing}. 
	 	\item \textit{Construct validity}. Test if the dimensions contribute to the overall evaluation of the questionnaire. Usually we have to identify the relationships between variables by means of Kaiser-Meyer-Olkin (KMO) test~\cite{KAISER1974}, which contrasts the partial correlations between the variables. The KMO statistic is a value between 0 and 1. A result below 0.5 is read as not being significant. It is also possible to confirm the validity of construct with the Barlett sphericity test~\cite{BARLETT1950}.
	 	\item \textit{Content validity}. It is the degree of comprehension of the questions. It also represents the quality of adjustment of the dimensions to the problem to be measured. There are two approaches used as test for validity: (a) the statistic~\cite{LAWSHE1975} that quantifies numerically the assessment data and (b) judges validation, in which a panel of experts close to the area and topics to be measured, give a detailed and independent assessments of the instrument~\cite{hyrkas2003validating}. Judges' validation is defined as a consensus among qualified persons who can issue evidence, judgments and valuations on the subject to be evaluated. Consensus methods for questionnaire validation includes the Delphi method and extensions (more information in Section~\ref{section31}). 
	 \end{itemize}
	\item \textit{Objectivity}. It is the degree to which this is or is not permeable to the influence of the biases and tendencies of the researcher or researchers who administer, qualify and interpret it.
\end{enumerate}	

In this paper we propose a method based on expert judgements~\cite{ding2002assessing}, as a technique to qualify the validity of content of a given questionnaire.

%

\subsection{Methodology and research question}\label{section22}

The communication possibilities offered by new technologies allow multitudes of new ways to teach and learn. Particularly in Higher Education students must be able to self-manage their learning processes and should be able to communicate effectively through the network. Current communication tools facilitate interaction and collaboration in virtual spaces. The use of \emph{flipped classroom} (FC) and \emph{m-learning} (mL) methodologies helps to overcome the distance between teachers and students and can improve the learning outcomes. 

With FC~\cite{BERGMANN2012flip} the learning activities are undertaken outside the classroom, through resources provided by the teacher, mainly videos, and activities provided to be solved inside the classroom and outside it (blending in-person with online learning moments), in a collaborative and meaningful way with the support of a facilitator (the teacher or tutor). FC main objective is to promote more active and responsible learning on the part of students~\cite{LAGE2000inverting}. 

One of the benefits of the mL methodology~\cite{JALDEMARK2017editorial} is to facilitate communication regardless of the time, devices and geographical location of the participants in the teaching-learning process~\cite{Zhang2015}. Thus, the m-learning methodology seeks to respond to the educational demand of the 21st century by providing advantages such as customizing learning experiences, achieving meaningful learning and developing professional skills.

Collaboration and virtual communication are fundamental aspects of b-learning because of the effect they have on learning and satisfaction~\cite{KIM2011763}. The use of a communication tool and an scenario of collaboration and communication between students, and between the teacher and the students, is the core of our piloting experience. The question that this research is intended to answer is: \emph{How can we promote virtual communication and satisfaction in Higher Education when FC and mL methodologies are applied?}

The theoretical model underlying the learning community is the \emph{Community of Inquire} (CoI) model~\cite{GARRISON199987,GARRISON201384}. According to this model, in the communication that takes place in a virtual community there are three styles of presence or core elements. These are: 
\begin{itemize}
	\item \textit{Cognitive Presence}: Through a series of phases, it allows the student to construct new educational experiences.
	\item \textit{Social Presence:} Develops interpersonal relationships through the media available in the learning environment.
	\item \textit{Teaching Presence:} Integrates the above elements through design, direct teaching and resource facilitation. It does not address directly the teacher/tutor, any one can play this role.
\end{itemize}	

Though there is research regarding data collection instruments for the CoI framework~\cite{Arbaugh08}, FC and mL methodologies are not implicit in the design. 

\subsection{Experience description}\label{section23}

The pilot experience puts in practice b-learning elements by applying combined methodologies of FC and mL in a Higher Education context, creating  synchronous and asynchronous virtual communications. Synchronous activities are planned as bidirectional communication channels with microblogging. Asynchronous activities follows a traditional scheme of delivering materials through a Learning Management System (LMS) such as Moodle. 

Resources of a course (documents, glossaries, quizzes, videos and other activities) are delivered to students using the \emph{Moodle} platform\footnote{Moodle \url{https://moodle.org}}. We selected as the tool for virtual communication the \emph{Telegram} app\footnote{Telegram \url{https://telegram.org}}, which allows synchronous and asynchronous communication, while it can be used from mobile devices, desktop multi-platform executable and the web. The communication scheme is given at Table~\ref{tbl:comType}.

In summary a total of eleven videos were created to teach some basic concepts of the subject (hardware classification, software licenses types, OS classification, etc.). Together with the videos there were individual and team activities that help to clarify and put in practice the concept learned. Feedback to students and solution to activities was given using Moodle platform and during the realization of structured follow-up sessions, that we called \emph{meetings}. 
Meetings were scheduled for each one of the Telegram group-teams, that we called \emph{planets}. Each planet chose the day and hour of the meeting to ensure maximum participation. Outside the meetings the teacher had to ignore the participants' conversations. In some cases, the students needed to catch the attention of the teacher (mentioning by the username) and she gave punctual answers. Most of the time, students were autonomous and free to communicate, provided they maintained a code of honor and good conduct. A questionnaire that measures the virtual communication and the satisfaction with this specific experience is needed. It should be a valid questionnaire.

\begin{table}
	\begin{center}
		\begin{adjustbox}{max width=\textwidth}
		\begin{tabular}{c|c|c|c}
			\hline \hline 
			Type of communication & Communication flow & Activity & Tool \\ 
			\hline 
			\hline 
			\multirow{3}{*}{Asynchronous}
			& Teacher $\Longrightarrow$  Students & Self-managed work & Videos at Moodle \\ 
			& Students $\Longrightarrow$  Teacher  & Set the meeting calendar & Google Doc with writing permission\\ 
			& Students $\Longleftrightarrow$  Students & Activities with the team & Telegram\\
			\hline 
			\multirow{2}{*}{Synchronous}
			& Students $\Longleftrightarrow$ Students & Informal team work & Telegram \\ 
			& Teacher $\Longleftrightarrow$ Students & Formal follow-up meetings & Telegram \\ 
			\hline \hline 
		\end{tabular}
	\end{adjustbox}
	\caption{Synchronous and asynchronous virtual communication experiences are designed.}
	\label{tbl:comType}
	\end{center}
\end{table}


This blended-learning experience was conducted in the first semester of the academic year of 2017-2018. The selected subject was \emph{Fundamentals of Software} (only with group E, teacher Rosana Montes) that it is conducted the first year of the Degree in Computer Engineering of the University of Granada. For most students, this is the first contact with higher education studies. To accommodate out the 70 students of group E in teams, we set eight \emph{planets} (The Earth, Mars, Venus,...) and let people freely distribute within, with a maximum of 10 participants per planet. 

A total of seven \emph{meetings} were scheduled for each of the \emph{planets}. The average duration of the meetings were 40 minutes approximately. Thousands of messages were produced between September and December 2017. It is not our aim to carry out an analysis of the content of the messages, but rather to measure the degree of appearance of each presence of the CoI model, along with the satisfaction with each presence and the general experience.

Given the popularity of FC and m-Learning, their researches have developed mostly independently. Is easy to find a questionnaire for each methodology separately, but it is most harder to find a questionnaire for a combined use of both methodologies. In the literature we only found a questionnaire specifically designed for a combined use of both methodologies~\cite{GARCIA201715}. This questionnaire tries to measure the satisfaction and communication in the underline CoI model. 

\subsection{Questionnaire definition}\label{section24}

Lets suppose that a given investigation covers $l$ different constructs or dimensions and that a questionnaire with $n$ items have been designed to evaluate those constructs. Each item is a text composed by two parts: the wording of the question and the scale to be used for the answer. Thus a questionnaire $Q$ is a succession of items $I = \{I_1,\dots,I_r,\dots,I_n\}\;(r = 1,\dots,n)$ grouped by dimensions $D$: 

$$
	Q~=~\{D_1,\dots,D_l\}~=~\{[I_1,I_i],[I_{i+1},I_j],[I_{j+1},I_u],\dots,[I_v,I_n]\}
$$

Table~\ref{tbl:dimBL} notes the questionnaire $Q_0$ from~\cite{GARCIA201715}. It has $l=7$ dimensions and $n=45$ items. This also poses a new challenge: experts in FC may not be experts in mL field. Particularly, we even needed experts in the CoI model. Thus, a validation method based on experts judgments should consider to use a very diverse expert panel and reflect the degree of expertise of each expert and for each dimension. 

\begin{table}[]
	\centering
	\resizebox{12.5 cm}{!} {
		\begin{tabular}{|
				>{\columncolor[HTML]{C0C0C0}}c |c|c|c|c|c|c|c|}
			\hline
			\multicolumn{8}{|c|}{\cellcolor[HTML]{C0C0C0}Questionnaire to evaluate a piloting experience in the Degree in Computer Engineering}                                                                                      \\ \hline
			{\color[HTML]{000000} Blocks}     & \multicolumn{3}{c|} {\begin{tabular}[c]{@{}c@{}}Virtual \\ Communication 
			\end{tabular}}       & \multicolumn{4}{c|} {\begin{tabular}[c]{@{}c@{}}Students' \\ Satisfaction	
			\end{tabular}}                       \\ \hline
			{\color[HTML]{000000} Dimensions} 
			& {\begin{tabular}[c]{@{}c@{}}Cognitive \\ Presence\end{tabular}} 
			& {\begin{tabular}[c]{@{}c@{}}Social \\ Presence\end{tabular}} 
			& {\begin{tabular}[c]{@{}c@{}}Teaching \\ Presence\end{tabular}} 
			& {\begin{tabular}[c]{@{}c@{}}Cognitive \\ Presence\end{tabular}} 
			& {\begin{tabular}[c]{@{}c@{}}Social \\ Presence\end{tabular}} 
			& {\begin{tabular}[c]{@{}c@{}}Teaching \\ Presence\end{tabular}} 
			& {\begin{tabular}[c]{@{}c@{}}General \\ Satisfaction\end{tabular}} \\ \hline
				{\color[HTML]{000000} Items} 
			& {\begin{tabular}[c]{@{}c@{}}$I_1$ - $I_8$ \end{tabular}} 
			& {\begin{tabular}[c]{@{}c@{}}$I_9$ - $I_{14}$  \end{tabular}} 
			& {\begin{tabular}[c]{@{}c@{}}$I_{15}$ - $I_{21}$  \end{tabular}} 
			& {\begin{tabular}[c]{@{}c@{}}$I_{22}$ - $I_{28}$  \end{tabular}} 
			& {\begin{tabular}[c]{@{}c@{}}$I_{29}$ - $I_{35}$  \end{tabular}} 
			& {\begin{tabular}[c]{@{}c@{}}$I_{36}$ - $I_{41}$  \end{tabular}} 
			& {\begin{tabular}[c]{@{}c@{}}$I_{42}$ - $I_{45}$  \end{tabular}} \\ \hline
			
		\end{tabular}
	}
	\caption{Blocks, Dimensions and Items corresponding to the questionnaire to evaluate Virtual Communication and Students' Satisfaction in FC and m-learning methodologies.}
	\label{tbl:dimBL}
\end{table}

\section{A 2-Tuple Fuzzy Linguistic Delphi method}\label{section3}

Our proposal is to define a 2-Tuple Fuzzy Linguistic Delphi (2TFLD) method to be used to test validation of construct of a given questionnaire. It its inspired in the Fuzzy Delphi method which is explained in Section~\ref{section31}. Preliminaries definitions for the underlying linguistic model are given in Section~\ref{section32}. Section~\ref{section33} specifies the general steps that should be conducted for the application of the 2TFLD method. Section~\ref{section34} defines the MEMCLDM problem behind the evaluation of each item of the questionnaire. Finally, Section~\ref{section35} explains the computation the consensus index that measures agreement between judges.

\subsection{The Delphi method} \label{section31}
The Delphi method was developed by Norman Delkay and Olaf Helmer in 1963~\cite{dalkey1963}. It is a iterative process used to collect and extract expert opinions using a series of questionnaires with interspersed feedback. Each version of the questionnaire is based on the previous iteration, ending when the expected consensus is reached. It has the characteristic of being anonymous, in order to avoid influence in the answers~\cite{dalkey1969}.

In the Delphi method, before a solution can be obtained it is necessary to ensure that a satisfactory level of consensus is reached. Consensus processes refers to how to reach the maximum degree of agreement between individuals or experts on the set of alternative solutions. It is usually coordinated by a person, called a moderator, who helps individuals or experts bring their opinions together~\cite{KACPRZYK199221,saint1994rules}.

To apply the Delphi method the researcher starts with a \textbf{preliminary phase} consisting of: (1) the identification of the problem and features, (2) the establishment of a coordination group that prepares the pilot questionnaire, and (3) the selection of the panel of experts that provide their opinion on the assessment of each item of the instrument in every iteration carried out. 

The selection of the expert panel is extremely important. Skjong and Wentworht~\cite{skjong2001expert} proposed the following selection criteria: (1) expertise in the area, (2) reputation in academia, (3) availability and motive, and (4) impartiality. According to Lynn~\cite{LYNN1986} a panel of three experts are required at a minimum.

Later it follows the \textbf{assessment phase}, which is the core phase of the method. It is conducted by the experts panel guided by the moderator. The following steps repeat in a finite loop: 

\begin{enumerate}[i]
	\item Disseminate the instrument to judges, independently.
	\item Sort, assess and compare the responses obtained in the first iteration.
	\item Modify the instrument items according to the judges' suggestions.
	\item Disseminate the new version to judges, independently.
	\item Feedback the judges on each iteration. Stop when responses show positive consistency or an acceptable consensus degree between judges has been achieved. 
\end{enumerate}	

Consensus is the agreement produced by consent between all members of a group or between several groups. To reach consensus, a figure of a moderator is key to gather judges' suggestions and assessments and to generate a new version of the instrument for the following iteration integrating judges' suggestions. Usually the moderator is the same researcher that has design the instrument. It is under his/her responsibility to accept or reject those suggestions, but in any case, the panel of experts have to assess the modified version of the questionnaire on each new iteration. 

As Ishikawa~\cite{ISHIKAWA1993max} pointed out, the classical Delphi method takes too much time and that represents high costs of application. The problem is that to reach an adequate degree of consensus it is required several iterations. It is important to mention that to speed up this Delphi technique, is usually acceptable to evaluate each item with a binary scale (reject or accept) with the corresponding loss of information and knowledge inherent in the expert panel. To avoid that, we propose to use many descriptive scales. 

In addition, expert assessment judgements may contain ambiguity because of the different interpretation each of them has of the instrument. In real situations, our personal interpretations are best reflected by using qualitative values, because words are close to human way of reasoning. 

Murray et al.~\cite{MURRAY1985} proposed the Fuzzy Delphi (FD) method as a combination of fuzzy sets and the Delphi method to solve some of the disadvantages of the classical Delphi method. The FD enables the interpretation of responses from a linguistic perspective~\cite{ZADEH1975}, by involving fuzzy numbers~\cite{ZADEH1965} as the representation of words, providing more reasonable results. These tools avoid confusion and allow a common understanding between expert opinions~\cite{NOORDERHAVEN1995strategic}.

To represent a linguistic variable without loss of information, different representation models have been used. Our proposal is to extend the FD by apply the 2-tuple fuzzy linguistic model~\cite{HERRERAYMARTINEZ2000} as a representation that deals positively with the aggregation processes in CW.

\subsection{The linguistic representation model}\label{section32}

To implement a CW based linguistic ME-MCLDM system, a model for linguistic data representation have to be chosen. It is our interest to provide a flexible \emph{2-tuple fuzzy linguistic delphi} method to be used by the scientific community: (1) as a tool for a researcher to validate a questionnaire and, (2) as an informative tool for the researcher that, as an expert, has to reach an appropriate degree of consensus with others. We foreseen to incorporate the following characteristics:

\begin{itemize}
\item
	The iterative nature of the Delphi technique force to understand the results of the previous iteration. The collective opinion computed by a MEMCLDM model should be a word which is easier to understand that statistical measures such as the standard deviation or the KMO values, because words are close to human way of reason. Linguistic outputs are obtained thought the use of the 2-tuple fuzzy linguistic model~\cite{HERRERAYMARTINEZ2000}. The better understanding of the collective opinion will favor consensus-reaching processes.
\item
	The expert can choose between several linguistic term sets the one that better suits his/her degree of expertise. Most of the times a questionnaire covers many different constructs and some constructs could be distant to the expert. For instance, when the research applies different methodologies by combination. In these situations a particular expert can have high confidence in some constructs and less in others. Nonetheless the expert evaluates the questionnaire entirety and not in some parts. We incorporate the idea of expert weights per dimension (noted as $W_{D_m}$ in Section~\ref{section34}). We also assume that if you have high knowledge in a particular field, it is better to have a richer set of terms. In this way, we allow the expert to modify his/her scale at any time. 
\end{itemize} 

Hereunder, we describe the details of each feature and how it is related to our computational linguistic model.

\subsubsection{The 2-Tuple linguistic computational model}\label{section32a}

A linguistic variable can take values only in a finite set values defined in the linguistic term set $S=\{s_0,\dots,s_g\}$, in which $g+1$ is called the cardinality of $S$ and usually is an odd number. The more terms in $S$ the greater the richness of the qualitative expression. But on the contrary, it also imposes hesitation to the expert and small set of terms are generally preferable. 

To deal with imprecision and vagueness, a linguistic term $s \in S$ is defined by a fuzzy number represented with triangular membership function uniformly distributed. Under this assumption it is guaranteed that the 2-tuple fuzzy linguistic representation model~\cite{HERRERAYMARTINEZ2000} based on symbolic translation is precise and effective, as it is a continuous representation of a linguistic term or word, and avoids loss of information in computational processes.

\begin{figure}
	\centering
	\includegraphics[height=0.20\textheight] {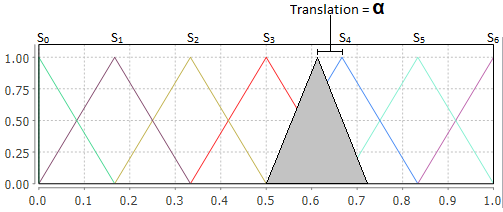}
	\caption{The value of $\alpha$ represents the translation of the membership function to the nearest term.}
	\label{fig:translation-alpha}
\end{figure}

\theoremstyle{definition}\label{def:tupla}
\begin{definition}~\cite{HERRERAYMARTINEZ2000}. A 2-tuple $(s_i,\alpha)$ (shown in Figure~\ref{fig:translation-alpha}) is a representation of the linguistic term $s_i \in S=\{ s_0, \dots, s_g\}$ for computations in CW processes.
\begin{enumerate}
\item 
	Let $s_i \in S$ be a linguistic term whose semantic is provided by a fuzzy membership function.
\item 
	Let $\alpha \in [-0.5,0.5)$ be the value of the \emph{symbolic translation} that indicates the translation of the fuzzy membership function representing the closest term when $s_i \in S$ does not exactly match the calculated linguistic information. 
\item
	A symbolic computation operates with the indexes of the linguistic terms and obtains a value $\beta \in [0, g]$.
\end{enumerate}
\end{definition}

\begin{obser}\label{obser:zero}
The transformation of a linguistic term into $s_i \in S$ into a linguistic 2-tuple is carried out by adding a zero as a symbolic translation to the linguistic term:
\end{obser}
$$
	s_i \in S \longrightarrow (s_i,0) 
$$

\begin{prop}Let $(s_i,\alpha)$ be a 2-tuple linguistic value. There is a function $\varDelta$ that translates a 2-tuple into a number $\beta \in [0,g]$:
\end{prop}

\begin{equation}\label{eq:2Beta}
\begin{array}{l}
	\varDelta^{-1}: S \times [-0.5,0.5) \rightarrow [0,g]\\
	\varDelta^{-1}(s_i,\alpha)~=~i + \alpha~=~\beta
\end{array}
\end{equation}

\begin{prop}Let $\beta \in [0,g]$ be the result of a symbolic computation. The equivalent 2-tuple is obtained my means of the function $\varDelta$ defined as:
\end{prop}

\begin{equation}\label{eq:2Tuple}
\begin{array}{l}
\Delta: [0,g] \rightarrow S \times [-0.5,0.5)\\
\Delta(\beta)~=~(s_i,\alpha), \;with\; 
	\left\{
	\begin{array}{l}
	    s_i \;\;\;i=round(\beta), \\
	   \alpha = \beta - i
	\end{array} 
	\right .
\end{array}
\end{equation}

where \emph{round} is a function that assigns the nearest integer $i \in \{0,1,\dots,g\}$. 

Information aggregation plays a key role in the decision solving scheme. The aggregation of a set of linguistic 2-tuple values must be also a linguistic 2-tuple that summarizes this set. Many linguistic aggregation operators of 2-tuples have been defined in the literature \cite{MARTINEZ2012, WAN2013, WEI2012,XU2011} to make linguistic information aggregation much easier and more flexible.

Let $x = \{(s_1,\alpha_1),\dots,(s_n,\alpha_n)\} = \{\beta_1,\dots,\beta_n\}$ be a set of linguistic values represented as 2-tuples, $W$ a  weighting vector ($\{w_i | i=1,\dots,n\}$), and $W'$ its normalized version such as $\sum_{i=1}^n w'_i = 1$. Thus the arithmetic weighed extended mean $\bar{x}^e$ is defined as:

\begin{equation}\label{eq:avgTuples}
	\bar{x}^e (x)=\Delta\left(\frac{\sum_{i=1}^n\Delta^{-1}(s_i,\alpha_i) \cdot w_i}
	{\sum_{i=1}^n w_i} \right)
		=\Delta\left(\frac{1}{n}\sum_{i=1}^n\beta_i w'_i \right).
\end{equation}

\subsubsection{Multi-granular linguistic information}\label{section32b}

The expert panel is allowed to individually select the linguistic term set that best suits the act of elicit linguistic information. This is a flexible approach to express opinions. Considering the problem to be addressed several linguistic term set are defined varying semantic and/or granularity. Then for any assessment, the given linguistic term could belong to any of the set term sets. To cope with general scenarios, we use $S^3$, $S^5$ and $S^7$ linguistic term sets with three, five and seven linguistic labels respectively. 

The 2-tuple fuzzy linguistic model has facilitated the processes of CW with multiple linguistic scales through mainly two approaches: (1) linguistic hierarchies, a linguistic structure to deal with different linguistic scales in a symbolic way, and (2) the fusion approach, which initially deals with the fuzzy semantics of the linguistic terms and obtains results expressed by 2-tuple linguistic values. 

A Linguistic Hierarchy (LH)~\cite{HERRERAYMARTINEZ2001} is the union of a set of linguistic term sets, symmetrically distributed with an odd granularity of uncertainty, $n(t)$, where $t$ is a valid level of the hierarchy. 

$$
\begin{array}{l}
  LH~=~\cup_t\;\;S^{n(t)}(t)\,,\;\;t \in \{1,\dots,h\}\;\;\;\;h=3\\
  S^{n(t)}(t)=\{s_0^{n(t)},\dots,s_{\delta_t}^{n(t)}\}\;\; where\;\;\delta_t=n(t)-1,\;\;\delta_t \in \mathbb{N}. 
\end{array}
$$

To make smooth transitions between successive levels, a term in $S^{n(t+1)}$ is the midpoint of each pair of terms belonging to the previous level $t$. Labels of this term set are known as former modal points. A set of former modal points of level $t$ is defined as:
$$
    FP_t~=~\{ fp_t^0, \dots, fp_t^{2\delta_t}\}
$$

The previous situation pose a limitation in $LH$ as $S^7$ cannot be obtained from level $2$ with $S^5$ ($S^5$ can be obtained from level $1$ with $S^3$). The solution is the use of Extended Linguistic Hierarchies (ELH)~\cite{Espinilla2011}, which can manage any term set without any limitation. The extended linguistic hierarchies consist of a set of linguistic term sets $S^{n(t)}(t)$ that corresponds to a level $t$, each one with a different granularity $n(t)$. 

A new linguistic term set in the ELH with $t^*=m+1$ keeps all the former modal points. 
$$
   n(t^*)~=~\left(\prod_{t=1}^h\,\delta_t\right) + 1 ~=~ \delta_{t^*} + 1 
$$

It is possible to simplify $n(t^*)$ with the computation of the Least Common Multiple (LCM) value of the granularities of the family of term sets defined in the ELH. In this way:
$$
	n(t^*)~=~LCM(\delta_1,\delta_2, \delta_3)+1~=~LCM(2,4,6)+1~=~13
$$

\begin{figure}[h]
\begin{center}
\includegraphics[width=8.64cm,height=8cm]{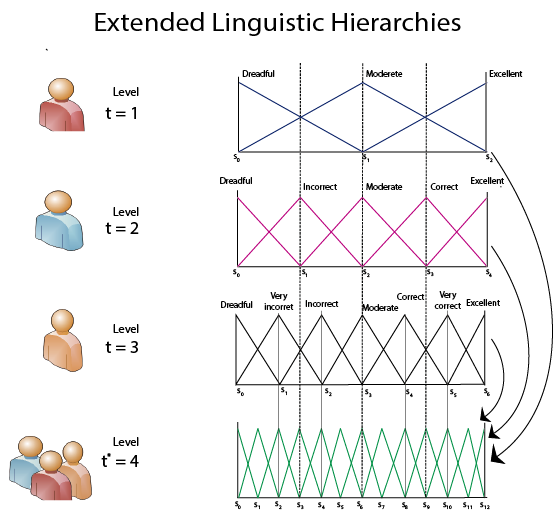}
\end{center}
\caption{Unification step translates linguistic values in $S^3$, $S^5$ or $S^7$ to level $t^*$, which in this case is $S^{13}$.}
\label{fig:ELH}
\end{figure}

The previous expression means that our computations are done under $S^{13}$, which is the bigger scale with common multiplier, as it is shown graphically at Figure~\ref{fig:ELH}. In a ELH, each formal model point $fp_{t^*}^i \in [0,1]$ is located at:
$$
  j~=~\frac{i \cdot \delta_{t^*}}{\delta_t} \to FP_t \subset FP_{t^*} \;\;\forall t = \{1,\dots,h\}
$$

The use of multiple linguistic scales simply adds a \emph{Unification} step in our decision solving scheme. The unification process translates the linguistic input data by means of a transformation function $TF_{t*}^t$ and converts a term $s_j \in S^{n(t)}$ into the equivalent term $s_k$ expressed in $S^{n(t^*)}$ with $t < t^*$. In this way computations are always conducted at level $t^*$, that keeps all the formal model points and is able to represent any value of any linguistic term set. Unification is guided by this transformation function: 

\begin{equation}\label{eq:transFunct}
   TF_{t^*}^t(s_j^{n(t)}, \alpha_j)~=~\Delta\left(\frac{\Delta^{-1}(s_j^{n(t)}, \alpha_j)\;\cdot\;(n(t^*)-1)}{n(t)-1}\right)~=~(s_k^{n(t^*)}, \alpha_k)
\end{equation}

\begin{example}\label{ex:1}
Opinions over an alternative regarding a criterion are elicited according to different term sets at $n(1), n(2), n(3)$. Unification is resolved as follows:
$$
\begin{array}{l}
	n(1)=3\;;(s_{1}^{3}, 0) \Rightarrow TF_{13}^3 = \Delta(\frac{1 \cdot 12}{2}) = (s_{6}^{13},0)\\
	n(2)=5\;;(s_{3}^{5}, 0) \Rightarrow TF_{13}^5 = \Delta(\frac{3 \cdot 12}{4}) = (s_{9}^{13},0)\\
	n(3)=7\;;(s_{4}^{7}, 0) \Rightarrow TF_{13}^7 = \Delta(\frac{4 \cdot 12}{6}) = (s_{8}^{13},0)\\
\end{array}
$$
\end{example}
 
\subsection{Stages of application of the 2-Tuple Fuzzy Linguistic Delphi method}\label{section33}

We propose to extend the FD method by addressing experts judgments with linguistic information, represented with ELH and the 2-tuple fuzzy computational model. The application of the 2TFLD method are those steps described in Section~\ref{section31}, here they are extended according to the notation used in the definition of a questionnaire in Section~\ref{section24}:

\begin{itemize}
	\item \textbf{Preliminary phase}:
	\begin{itemize}
		\item The research team define formally the problem to be evaluated and design the items of a questionnaire $Q_0$.
		\item The research team select and invite experts in the area to the panel of experts $J$. Optionally the research team assign to each expert a weight value with respect to $D$, the dimensions of $Q$. 
		\item The research team select a member to act as a moderator.
		\item The research team select a family of $h$ linguistic term sets, with their semantics. We propose to use $h=3$ with $S^3$, $S^5$ and $S^7$.
	\end{itemize}
	\item \textbf{Assessment phase}:
	\begin{itemize}
		\item Moderator starts the 2TFLD method by distributing $Q_0$ to each expert. 
		\item The expert choose a scale $S^{n(t)}$ to assess the questionnaire on the initial iteration $Q_0$, the next iteration $Q_{1}$, and so on ($Q_\iota$ with $\iota \leq max\_iterations$). 
		\item On each iteration moderator use a DSS tool (detailed in Section~\ref{section4}) to assist the validation by consensus (described in Section~\ref{section35}). Moderator can vary parameter $\epsilon$ or the \emph{satisfactory reliance level} under the same input data and gather a global view of the consensus at $Q_\iota$.
		\item Moderator set a new version of the questionnaire $Q_{\iota+1}$ that incorporates open suggestions. Then repeats the assessment phase with this new questionnaire.
		\item The procedure stop at a maximum number of iterations or when a satisfying level of consensus is achieved. 	
	\end{itemize}
	\item \textbf{Exploitation phase}:
	\begin{itemize}
		\item When core processes finalized, the research team have a complete overview of the evaluation of the questionnaire, as a whole, grouped by criteria and individually for each item, supported by an online tool.   
		\item Last version of the questionnaire might be used in a piloting experience to conduct statistical analyses such as: Cronbach's alpha, KMO index or Berlett's sphericity to corroborate if \emph{reliability}, \emph{validity} or \emph{objectivity} are met.
		\item If there is enough statistical confidence, the questionnaire can be applied in a real study.
	\end{itemize}
\end{itemize}	

\begin{figure}[h]
\begin{center}
\includegraphics[width=13.66cm,height=7.97cm]{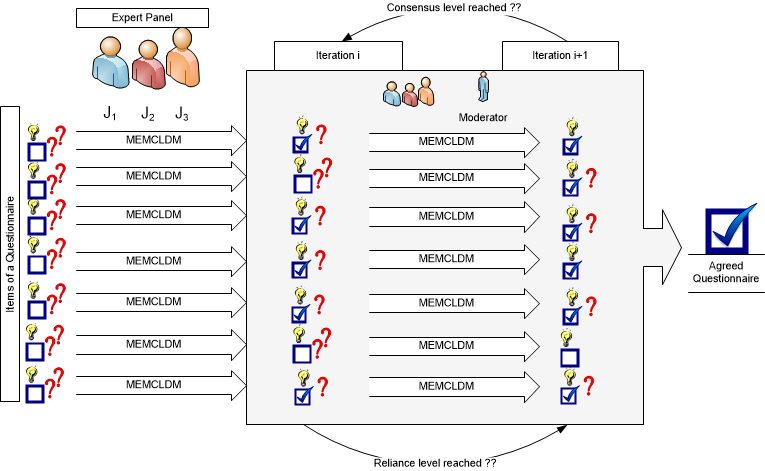}
\end{center}
\caption{The proposed 2TFLD method solves several MEMCLDM problems that are repeated trough iterations till a consensus level is reached for each item.}
\label{fig:general}
\end{figure}

Graphically the general CW scheme for our proposal is shown in Figure~\ref{fig:general}. As the classic Delphi method it is an iterative problem guided by a moderator figure. In our approach an item of a questionnaire $I_r$ is accepted by consensus or rejected as a result of a MEMCLDM problem.

\subsection{The MultiExpert-MultiCriteria Linguistic DM problem}\label{section34}

A questionnaire $Q_\iota$ is a succession of items $I = \{I_1,\dots,I_r,\dots,I_n\}\;(r = 1,\dots,n)$ grouped by $l$ dimensions, $D$, thus: 
$$
	Q_\iota~=~\{D_1,\dots,D_l\}~=~\{[I_1,I_i],[I_{i+1},I_j],[I_{j+1},I_u],\dots,[I_v,I_n]\}
$$
We consider that to test a questionnaire of size $n$, we have in fact to solve $n$ instances of the same MEMCLDM problem. This problem is defined considering the following:

\begin{itemize}
\item
	A single alternative is evaluated: the item $I_r$, its wording and its answering scale.
	
\item
	Let $J=\{J_1,\dots,J_p\} (i=1,\dots,p)$ be the expert panel involved in the validation of a questionnaire $Q_\iota$.
	
\item 
	Opinion is elicit in the form of a single label $s^{n(t)} \in S^{n(t)}$ with $S^{n(t)} \subset ELH$. According to the linguistic representation given in Section~\ref{section32}, after the processes of translation and unification (with Eq.(\ref{eq:2Tuple}) and Eq.(\ref{eq:transFunct})), we get a 2-tuple linguistic value expressed in a term set of level $t^*$ with $\delta_{t^*}=12$.
	
\item
	Judges can have different degrees of expertise and for that reason they are rated according the expertise over each dimension by the research team. Let suppose that $I_r \in D_m$ with $m = \{1,\dots,l\}$. Then $W_{D_m} =~\{w_{1D_m},\dots, w_{pD_m}\}$ is a normalized vector of size $p$ that it is used to give more relevance to the opinions of those judges with high weights.
		
\item	
	Each item $I_r$ is assessed according to $C=~\{C_1,\dots,C_q\}$, a set of $q=4$ linguistic criteria:
   \begin{itemize}
	\item 
		\emph{Clarity}. Is the the quality of being clear. In particular, the quality of being coherent and intelligible.
	\item 
		\emph{Writing}. It measures the skill of writing, that it is, the degree of proofreading in writing.
	\item 
		\emph{Presence}. It measures the pertinence of the item into its dimension. Sometimes the item is well formed but it is placed in the wrong dimension.
	\item 
		\emph{Answering scale}. According to the wording of the item, it assess the rightness of the answering scale.
	\end{itemize}

\item 
	Each item $I_r$ is also assessed according to a numerical property which characterizes the global \emph{relevance} of the item. This represents the importance or utility of the item in the questionnaire for the given research hypothesis. Expert $J_i$ rates this property with a number $w_i^r \in [0,1]$ to be used as the item weight in the processes of computing the linguistic result. We note $R$ the array of values given by each expert. 
\end{itemize}

The 2TFLD method is an iterative and dynamic process aimed at achieving a high degree of agreement before making the decision that solves. It uses linguistic and numerical data which is managed as individual assessments with respect to item $I_r$. So for each item we compute: (1) the collective group opinion $Y^r$ by the aggregation of all individual opinions of experts with respect to every criteria; (2) the consensus index $CI_r$ is computed for each item by considering all the criteria; and (3) the reliance index $RI_r$ to determine if the item is valid. The previous procedure is repeated for all the elements of $I$, resulting in a new information for the overall questionnaire, that we call Questionnaire Score $QS \in S^7$. 

\begin{figure}[h!]
\begin{center}
\includegraphics[width=13.66cm,height=4.21cm]{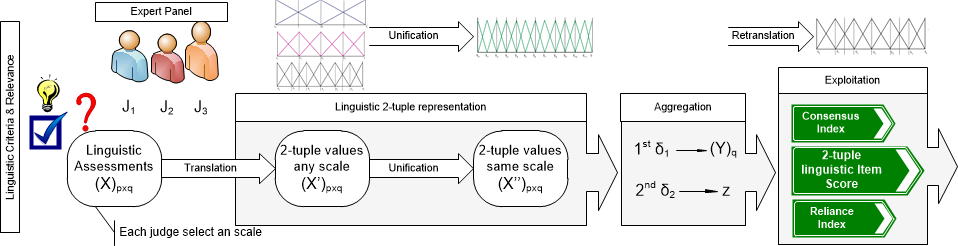}
\end{center}
\caption{The solution of a MEMCLDM through successive phases is	the qualification of a item of the questionnaire.}
\label{fig:single}
\end{figure}

Traditionally, the selection process for reaching a solution for a LDM problem after the definition of the problem, perform two main phases~\cite{Herrera2009}: (1) aggregation in which experts opinions are combined by using an aggregation operator, and (2) exploitation, that uses a selection criterion to obtain an alternative or a subset of alternatives as the solution to the problem. However, our proposal does not deal with different alternatives and it handles flexible ways of providing linguistic information. Because of that, we extend the classic processes to solve a MEMCLDM problem. The scheme is shown in Figure~\ref{fig:single}.

The following computational processes are better detailed:
\begin{itemize}

	\item \emph{Gathering phase}. 
An opinion is a single label represented with the 2-tuple fuzzy linguistic model according to an extended linguistic hierarchy. For simplicity let us compact the notation by using $x_{ij} \in S^{n(t)}$ for a given level $t$ selected by the judge. A linguistic information $x_{ij}^r$ is given by judge $J_i$ regarding criterion $C_j$ when s/he rates item $I_r$. We denote $X_r~=~(x_{ij}^r)_{p \times q}$ the linguistic assessment matrix. Experts also express his/her opinion regarding the relevance of $I_r$ with numerical values. 

\begin{table}[ht!]
  \centering
  $I_r \longrightarrow$
  \begin{tabular}{c|cccc|c} 
	\hline \hline 
   	 &$C_1$ & $C_2$ &$C_3$ &$C_4$ & $R$\\
	\hline
	\hline $J_1$ 	& $x_{11}^{r}$ 	& $x_{12}^{r}$	& $\dots$	 & $x_{1q}^{r}$ 	& $w_1^r$ \\
	\hline $\vdots$ 	& $\vdots$ 	& $\vdots$	& $\vdots$ & $\vdots$  	& $\vdots$ \\
	\hline $J_p$ 	& $x_{p1}^{r}$ 	& $x_{p2}^{r}$ 	& $\dots$	 & $x_{pq}^{r}$	& $w_p^r$ \\
	\hline \hline 
	\end{tabular} 
  \caption{Full assessment matrix for item $I_r$.}
  \label{tab:sampleMatrix}
\end{table}

	\item \emph{2-tuple transformation phase}. 
Linguistic values are represented using the 2-tuple fuzzy linguistic computational approach, so anly label $x_{ij}^r$ is translated to $(x_{ij}^r,0)$. Following an standard scheme of CW processes~\cite{Yager2004}, the translation would also imply the re-translation phase (using Eq.~(\ref{eq:2Tuple})). Let us note $X'_r~=~(x_{ij}^{'r})_{p \times q}$ the matrix of linguistic 2-tuples values. 

	\item \emph{Unification phase}. 
This phase unifies linguistic data by expressing all of them in the same linguistic term set $S^{n(t*)}$, which is the one with the highest cardinal. We use function $TF_{t^*}^t$ by applying Eq.~(\ref{eq:transFunct}). We denote $X^{''}_r~=~(x_{ij}^{''r})_{p \times q}$ the assessment matrix with 2-tuple linguistic terms in $S^{13}$.

	\item \emph{Double aggregation phase}. 
We combine opinions so that we anonymize the assessments and obtain a valuation of the item for each criterion. This information is be very useful for the moderator to provide feedback to the full panel of experts. Aggregation is carried out with the arithmetic weighed extended mean operator, $\bar{x}^e(W_{D_m})$ of Eq.~(\ref{eq:avgTuples}), with consideration of the expert weights $W_{D_m}$ with respect item $I_r \in D_m$ and for each judge $J_i \in J$. For simplicity in the notation we use $\delta_1=\bar{x}^e(W_{D_m})$ hereon. 

We apply $\delta_1$ to the linguistic information $X^{''}_r$ to get vector $Y_r~=~(y_{j}^{r})_q$. We also compute the item average relevance $W^r = \frac{1}{p}\sum_{i=1}^p\;w_j \cdot w_{iD_m}$. Thus we obtain the following data:

\begin{table}[ht]
  \centering
  \begin{tabular}{c|cccc|c} 
	\hline \hline 
   	 &$C_1$ & $C_2$ &$C_3$ &$C_4$ &$R$ \\
	\hline
	\hline $\delta_1(X^{''}_r)=Y_r$ 	& $y_{1}^{r}$ 	& $y_{2}^{r}$	& $\dots$	 & $y_{q}^{r}$ & $W^r$ \\
	\hline \hline 
  \end{tabular} 
  \caption{Vector notation for the first aggregation.}
  \label{tab:sampleMatrix}
\end{table}

The next step considers a second aggregation only over the 2-tuple linguistic values $y_j^r$, leaving apart $W_r$. In this way, we compute the collective overall opinion for item $I_r$ by aggregating criteria values with the aggregation operator $\bar{x}^e(V)=\delta_2$ instantiated with a vector of uniform weights ($V=\{v_j=1/q \;\;|\;\; j=1,\dots,q\}$). 

\begin{equation}\label{eq:collective}
	\delta_2(Y_r)=Z_r~=~(s_z^r,\alpha_z^r) \;\;\;with\;\;\;s_z^r \in S^{n(t^*)}
\end{equation}

	\item \emph{Exploitation phase}. 
We are able to provide two types of results with the information computed in the aggregation phase:
    \begin{itemize}
	\item 
	Individual results for item $I_r$. The value $(s_z^r,\alpha_z^r)$ from Eq.~(\ref{eq:collective}) is the main output to the expert panel. Nevertheless, we also have to compute $CI$ and $RI$ for the knowledge of the moderator. Boolean values $CS$ and $RS$ could be shared with the expert panel to better understand the circumstances for a new iteration.
	For the sake of understanding, given that experts $J$ expressed themselves over $S^3$ or $S^5$ or $S^7$ we apply a re-translation from level $t^*$ to level $t=h=3$. The linguistic output is known as the Item Score $IS_r$ with $s_r \in \{Dreadful, Incorrect, Moderate, Correct, Very\,correct,\\Excellent\}$.
\begin{equation}\label{eq:ToS7}	
	IS_r~=~TF_{t^*}^{3}(s_z^r,\alpha_z^r)~=~(s_r,\alpha_r)
\end{equation}
	\item 
	Collective results for questionnaire $Q$. A questionnaire is in fact a set of $n$ items, each one with a linguistic score $IS_r$ and a collective relevance opinion $W^r$. To provide more information, 2TFLD uses relevance values as weights in a third aggregation step $\delta_3=\bar{x}^e(CW)$ with $CW = \frac{1}{n} \sum_{r=1}^n W^r$, to compute global scores for $Q$ as 2-tuple linguistic values in $S^7$:
	\begin{itemize}
		\item Collective Clarity: $CC = \delta_3(Y_{C_1})$ with $Y_{C_1} = \{y_1^1, \dots,y_1^n\}$.
		\item Collective Writing: $CW = \delta_3(Y_{C_2})$ with $Y_{C_2} = \{y_2^1, \dots,y_2^n\}$.
		\item Collective Presence: $CP = \delta_3(Y_{C_3})$ with $Y_{C_3} = \{y_3^1, \dots,y_3^n\}$.
		\item Collective Answering Scale: $CAS = \delta_3(Y_{C_4})$ with $Y_{C_4} = \{y_4^1, \dots,y_4^n\}$.
		\item Questionnaire Score $QS = \delta_3(CIS)$ with $CIS = \{IS_1, \dots,IS_n\}$.
	\end{itemize}	
	 
   \end{itemize}

\end{itemize}

\subsection{Validation by Consensus}\label{section35}

The consensus process tries to achieve the maximum degree of consensus possible among the opinions of individuals or experts. The degree of consensus is calculated at each stage. If the grade is satisfactory, then the questionnaire is positively tested. Conversely, if the degree of consensus measured is not satisfactory, then individuals or experts are encouraged to modify their views in order to increase proximity in their approaches. In this way, we set a Decision Making process dynamic and iterative in which individuals or experts change their opinions until their approaches to the solution are sufficiently close, at which point, consensus is reached. 

To compute the degree of consensus we need to measure the difference between individuals and collective opinions, which is in fact a measure of error. Thus it is desirable to get differences values close to zero, meaning that opinions of members of a group are similar. 

From Section~\ref{section34} we have $p$ judges to elicit $n$ linguistic decision matrices $X'_r$. We also have $n$ 2-tuple linguistic collective opinion $Y^r$ with respect to a criteria set of $q$ elements. Finally for each item, the 2-tuple linguistic output value is $(s_z^r, \alpha_z^r)$. Also per item, we compute a separation measure $\rho \in [0,\infty]$ for each judge $J_i$:

\begin{equation}\label{eq:separation}
	\rho_i = \sqrt{ \sum_{j=1}^q \left( \Delta^{-1}(x'_{ij}) - \Delta^{-1}(y_{j}) \right) ^2}\;,\;\;i~=~1,\dots,p
\end{equation}

When an expert sees his/her $\rho$ value high, it means that in general s/he has not very similar opinions to the collective.

\theoremstyle{definition}\label{def:consensus}
\begin{definition}
Let Consensus Index $CI_r \in [0,1]$ be our measure of consensus between experts regarding item $I_r$. We consider that the information collected from the judges could be influenced by vectors of normalized weights $\{W_{D_1},\dots,W_{D_l}\}$ that represent the expertise degrees defined for each dimension of the questionnaire, were $v_i \in W_{D_m}$ if $I_r \in D_m$. Let $CS_r$ be a boolean value that takes \emph{true} if there is consensus, that is when $CI_r \ge 0.5$ or \emph{false} in other case. According to these assumptions, the consensus index is defined as:

\begin{equation}\label{eq:CI}
	CI_r~=~ 1 - \frac{\sum_{i=1}^p \rho_i \cdot v_i}{\delta_{t^*}}\\
\end{equation}
\end{definition}

In our opinion, consensus processes need to be flexible and adjustable by the moderator. If there is a perception of when the consensual solution is good, we can graduate the level of demand of consensus processes. That is why we use a parameter called \emph{satisfactory reliance level} $\epsilon \in [0,1]$ that determines that consensus can be reached in a smaller number of iterations, when $\epsilon$ is close to zero. If $\epsilon$ is close to one, it will be harder for the expert with high $\rho_i$ to narrow the gap with the group. Thus, this parameter represents when the solution is acceptable to the moderator. 

\begin{definition}\label{def:reliance}
We define the Reliance Index $RI_r \in [0,1]$ of an item $I_r$ as follows:

\begin{equation}\label{eq:RI}
	RI_r~=~ \sum_{j=1}^q u_j \;\;\;\; where\;\;u_j =
	\left\lbrace 
  	\begin{array}{ll}
     		1/q & if~ \Delta^{-1}(y_j) \ge \delta_{t^*} \epsilon\;,\\
     		0 & else.
  	\end{array} \right. 
\end{equation}
\end{definition}

In this sense, $RS_r$ is the boolean value that takes \emph{true} when $RI_r \ge \epsilon$ or \emph{false} in other case. Note that in assessment phase, CW processes are done at level $t^*$. 

\begin{example}\label{ex:2}
In this example moderator set $\epsilon=0.6$. Consider that opinions regarding $I_1$ from $J=\{J_1,J_2,J_3\}$ are the same that in Example~\ref{ex:1}, thus $X' = \{(s_6^{13},0), (s_9^{13},0), (s_8^{13},0)\}$ with $W_{D_1}=(0,2,0.6,0.2)^T$. For simplicity, we assume single criteria ($q=1$) and thus $(s_z,\alpha_z)=(s_8^{13}, 0.2)$. 

By using Eq.(\ref{eq:separation}) we have $\rho = \{ 2.2, 0.8, 0.2\}$ that reflects that $J_1$ has the most distant opinion from the solution, nevertheless $v_1=0.2$ is low, and total consensus is positive with $CI_1=0.92$ by applying Eq.(\ref{eq:CI}). The previous values set $CS_1=true$ and $RS_1=true$ because $RI_1=1$. 

Yet, if we change the model parameter to $\epsilon=0.8$ the overall situation changes, given that the inequality $\Delta^{-1}(s_8^{13}, 0.2) \ngeq 9.6$ from Eq.(\ref{eq:RI}). As a result, it sets $RI_1=0$ and $RS_1=false$. In this situation, the moderator needs another round of assessments to improve consensus and reliance levels.

\end{example}

\section{A web tool to apply the 2-Tuple Fuzzy Linguistic Delphi method}\label{section4}

People taking decisions should be provided with tools that assist them to make that decisions. This kind of software is know as Decision Support Systems (DSS). This work contributes with an online tool that implements the 2TFLD method presented in Section~\ref{section3}. In the following sections we describe its objectives (Section~\ref{section41}), key points of its functionality (Section~\ref{section42}) and the input file format (Section~\ref{section43}). 

\subsection{The 2-tuple-fuzzy-delphi DSS}\label{section41}

There are some online solutions for the application of the Delphi method: some are free as Delphi2\footnote{Delphi2 \url{http://armstrong.wharton.upenn.edu/delphi2/}}, and some are commercial software as Mesydel\footnote{Mesydel \url{http://www.spiral.ulg.ac.be/en/tools/online-delphi-mesydel/}} and Surveylet\footnote{Surveylet \url{https://calibrum.com}}. These tools are not licensed for adaptation or modification, so it is difficult to put in practice suitable linguistic representation models or solution schemes of Computing with Words. This indicates that the Delphi method is active in the research community and that there is an opportunity to assist the iterative processes reducing the cost of applying the method. 


The proposed \textbf{2-tuple-fuzzy-delphi} online tool purpose is to be a facility for the moderator to guide him/her in the task of reaching a consensual questionnaire. On each iteration, moderator imports original assessments from $J$ to visualize, as a whole, the individual and collective linguistic scores extracted from the solution of the MEMCLDM problem and the overall consensus. If reliance and consensus levels are also satisfiable the questionnaire can be used for piloting. Otherwise with the assistance of this tool, feedback is given to the expert panel.

This software is available at \url{https://sci2s.ugr.es/2tuple-fuzzy-delphi} for public use. Its code source is available under GNU GPL v3 license at GitHub repository \url{https://github.com/NoeZermeno/2tuple-fuzzy-delphi}.

\subsection{Key features of 2-tuple-fuzzy-delphi DSS}\label{section42}

After importing the data, the user has a tool to visualize the linguistic, numeric and boolean results of the application of the 2TFLD method. In addition, these features are accessible with the tool:
\begin{itemize}
	\item \emph{Filtering:} The user can list content using a selector to display any column data. Filtering options consider: Collective Clarity, Collective Writing, Collective Presence, Collective Answering Scale, Average Relevance, Consensus, All information. By default All information is select to have a full overview of all the information obtained through the analysis of judges' responses.
	\item \emph{Data simulation:} The user can affect the solution of the model by varying with a slider parameter $\epsilon$. By changing the satisfiable reliance parameter, moderator sets different acceptable level of consensus under the same input data. 	
	\item \emph{Searching:} There is a text searching tool which makes it easy for us to locate an item and focus on its scores.
	\item \emph{Trimming:} It may happens that the expert panel advice a reduction in the number of items. Then we have to answer this question: \emph{which elements should be removed to address my desirable number of items?}. We provide assistance to this task by implementing a trim operator. A radio button input scale considers $S^7$ linguistic terms and ranges from $s_0$ to $s_6$. It is set by default at $s_0$ (0 items trimmed). Ongoing with a interaction to the right, the user sees that some items hide and a label informs of the number of items trimmed. 
	\item \emph{Sorting:} Tabular data is better handled when the user is able to select by clicking, the criteria of sorting from A to Z with the first click, and from Z to A with a successive click under the same column.
\end{itemize}

\subsection{2-tuple-fuzzy-delphi DSS input file format}\label{section43}
Here we describe the details of the CSV format, the input file data used on each iteration of the 2TFLD method. The exploitation of a questionnaire, through its use as online surveys, could be supported by some well known online services, such as Google Forms\footnote{Google Forms \url{https://www.google.es/intl/en/forms/about/}}, Monkey Survey\footnote{Monkey Survey \url{https://www.surveymonkey.com}} or Lime Survey\footnote{Lime Survey \url{https://www.limesurvey.org}}. For that reason, we have developed a free and open licensed tool that copes with data exported from those to online services. A CSV file is also easily get from spreadsheets desktop solutions if the researcher does not use the previous services. In this way our tool is able to nourish data from external sources that are highly available.

Guided by how Google Forms exports as spreadsheets its content data, we have defined our input format. At the same time, the moderator can use this service solution to share the current version of the questionnaire with each judge of the expert panel. After the first round is complete, the moderator starts using our \emph{2-tuple-fuzzy-delphi} DSS. Google Forms collect all responses in a \emph{.gsheet} file that can be downloaded in OpenOffice format or in Microsoft Office format. Either case, particular sheets or pages can be exported to CSV file format individually to be used as our input data. It is be desirable to add some optional sheets to complement the gathered responses sheet. 

Generally for each round we can use up to three type of sheets. We suggest the following wording to be used as the sheet names, where $X$ represents the number of the current iteration. The following is a description of the content to be stored on each sheet:

\begin{itemize}
\item 
	\emph{Round$X$Description}. This data corresponds to the text description per item. Optionally the first row correspond to the header name. In that case, $n$ is the number of lines read minus 1. Content type in this case is: \emph{description}. This import is not mandatory as a generic text would be used in case of absence.

\item 
	\emph{Round$X$Dimensions}. Here data associates each judge's expertise with questionnaires dimensions. At the same time, associates items ranges with dimensions. Content type in this case is: \emph{dimensions}. The number of lines (minus one if headers are enabled) is $l$, the number of dimensions. This import is not mandatory as uniform weights would be used in case of absence, also eliminating the need to know which dimension an item belongs to. Table~\ref{tab:dimMatrix} describes the structure of this type of data.
	
\item
	\emph{Round$X$Responses}. It contains the MEMCLDM problem data thus it is mandatory. Content type in this case is: \emph{responses}. According to the number of rows an columns parsed, we compute the number of experts $p$ and the number of items $n$ respectively. Optionally the first row correspond to the header. A well formed CSV file complies with the  structure represented at Table~\ref{tab:sampleMatrix}.
\end{itemize}

\begin{table}[ht]
  \centering
  \begin{tabular}{c|cc|cccc} 
	\hline \hline 
	Dimension & Begin & End & $J_1$ & $J_2$ & \dots & $J_p$\\
	\hline \hline 
	$D_1$ & $I_1$ & $I_i$ &  $w_{1D_1}$  &  $w_{2D_1}$ & $\dots$ & $w_{pD_1}$\\
	\hline 
	$D_2$ & $I_{i+1}$ & $I_j$ & $w_{1D_2}$ & $w_{2D_2}$ & $\dots$ &  $w_{pD_2}$\\
	\hline $\vdots$	& $\vdots$ & $\vdots$  & $\vdots$ & $\vdots$ & $\dots$ & $\vdots$ \\
	\hline 
	$D_l$ & $I_v$  & $I_n$ & $w_{1D_l}$ & $w_{2D_l}$ & $\dots$ &  $w_{pD_l}$\\	

	\hline \hline 
  \end{tabular} 
  \caption{Structure of the data grid: subdivision of items into dimensions and expert weights per dimensions of the questionnaire. Header names are optional.}
  \label{tab:dimMatrix}
\end{table}

\begin{table}[ht]
  \centering
  \begin{tabular}{cc|ccccc|c|ccccc} 
	\hline \hline 
   	 Judge & Level &$C_1$ & $C_2$ &$C_3$ &$C_4$ & $R$  & $\dots$ &$C_1$ & $C_2$ &$C_3$ &$C_4$ & $R$\\
	\hline
	\hline $J_1$ & $n(t)_{J_1}$ & $x_{11}^{1}$ & $x_{12}^{1}$	& $\dots$	 & $x_{1q}^{1}$ 	& $w_1^1$ 
		& $\dots$ & $x_{11}^{n}$ & $x_{12}^{n}$	& $\dots$	 & $x_{1q}^{n}$ 	& $w_1^n$ 
	\\
	\hline $J_2$ & $n(t)_{J_2}$ & $x_{21}^{1}$ 	& $x_{12}^{1}$	& $\dots$	 & $x_{2q}^{1}$ 	& $w_2^1$ 
		& $\dots$ & $x_{21}^{n}$ & $x_{22}^{n}$	& $\dots$	 & $x_{2q}^{n}$ 	& $w_2^n$ 	
	\\	
	\hline $\vdots$ 	& $\vdots$ 	& $\vdots$	& $\vdots$ & $\dots$  	& $\vdots$ & $\vdots$  
		& $\vdots$ 	& $\vdots$	& $\vdots$ & $\dots$  	& $\vdots$ & $\vdots$ 
	\\
	\hline $J_p$ & $n(t)_{J_p}$ 	& $x_{p1}^{1}$ 	& $x_{p2}^{1}$ 	& $\dots$	 & $x_{pq}^{1}$	& $w_p^1$ 
		& $\dots$ & $x_{p1}^{n}$ & $x_{p2}^{n}$	& $\dots$	 & $x_{pq}^{n}$ 	& $w_p^n$ 
	\\
	\hline \hline 
  \end{tabular} 
  \caption{Structure for $Q$ data grid is similar to Google Form spreadsheets responses grid. Header names are optional.}
  \label{tab:sampleMatrix}
\end{table}

\begin{figure}[h]
\begin{center}
\includegraphics[width=13.66cm,height=6.9cm]{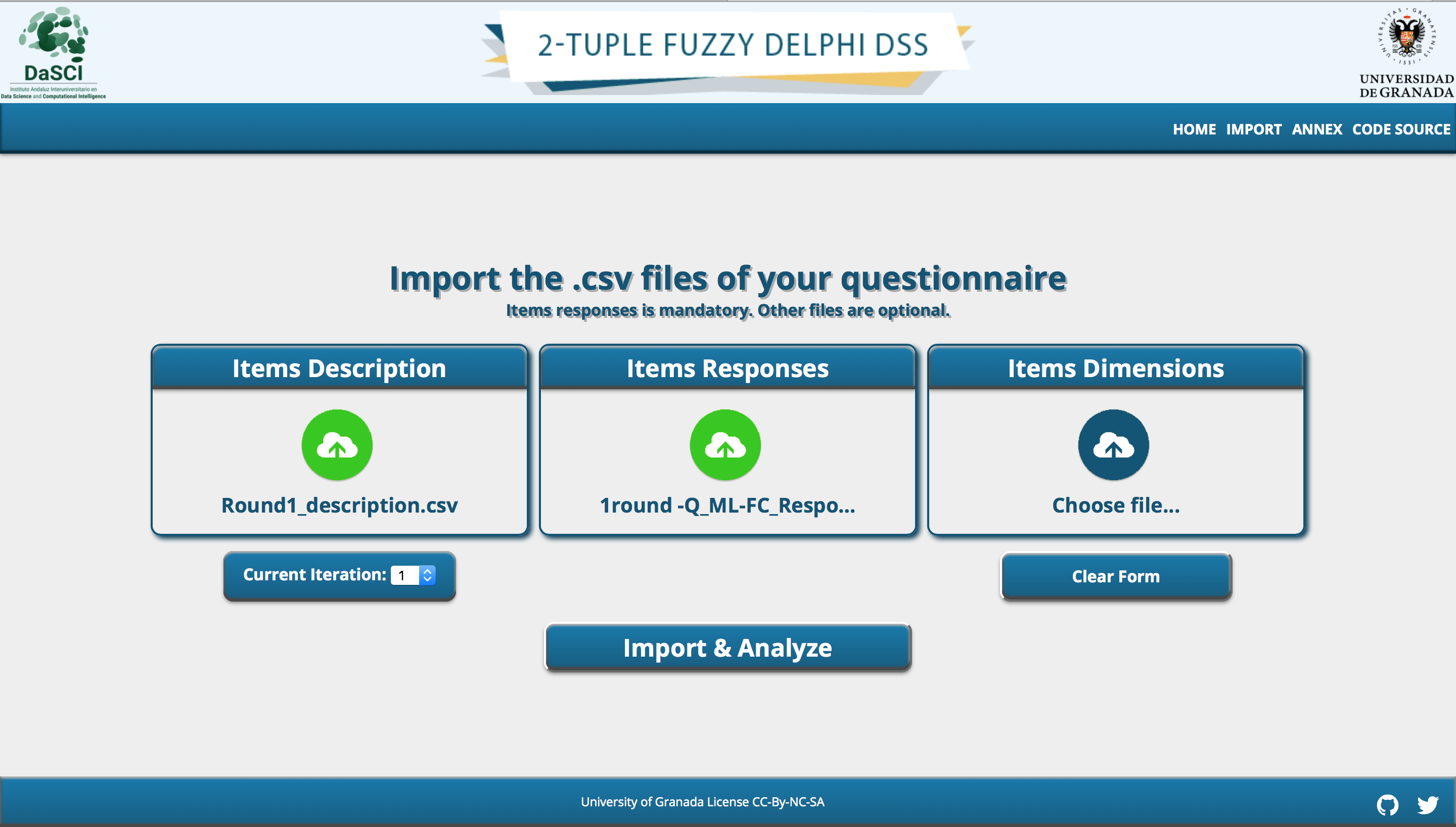}
\end{center}
\caption{We can import separately the description of the questionnaire and the assessments of the expert panel for each round.}
\label{fig:import}
\end{figure}

Each sheet separately exports as CSV file which can be imported in our site. This setting is shown in Figure~\ref{fig:import}. We require the following information: number of the current round, content type and file path for data (Responses data is the minimum required). 

\section{Case Study: content validity of a questionnaire for b-learning}\label{section5}

We have conducted a blended learning experience in education that put in use the flipped classroom and the mobile learning methodologies in combination. As we introduced in Section~\ref{section2}, we want to evaluate the experience and for that reason a questionnaire of $45$ items have been selected. We need to ensure content validity of the questionnaire by applying our 2TFLD proposal.

We configured our own committee of $p=9$ experts. Table~\ref{tab:dimBL} describes the expertise of each judge according the structure of the questionnaire and how items are grouped in $7$ dimensions. This is the content of the \emph{dimensions} sheet. 

We acted as moderators and used the \emph{2tuple-fuzzy-delphi} DSS. In the first iteration, we got a questionnaire score $QS_1 = (s_5, -0.389)$ or \emph{Very correct} (details are given in Section~\ref{section51}), along with a description of the changes applied to the questionnaire. After the second iteration $QS_2 = (s_6, -0.442)$ or \emph{Excellent} (details are given in Section~\ref{section52}). We achieved a consensual version, nevertheless the expert panel suggested a reduction in the number of items. The use of the trimming tool is described in Section~\ref{section53}. 

\begin{table}[ht]
   \centering
   \resizebox{14cm}{!} {
	\begin{tabular}{c|cc|ccccccccc}
		\hline \hline
		Dimension & Begin & End & $J_1$ & $J_2$ & $J_3$ & $J_4$ & $J_5$ & $J_6$ & $J_7$ & $J_8$ & $J_9$ \\ 
		\hline \hline
		$D_1$ & $1$ & $8$ & $0.118$ & $0.093$ & $0.087$ & $0.124$ & $0.112$ & $0.124$ & $0.112$ & $0.124$ & $0.106$ \\ 
		\hline 
		$D_2$ & $9$ & $14$ & $0.125$ & $0.094$ & $0.088$ & $0.119$ & $0.113$ & $0.113$ & $0.113$ & $0.125$ & $0.113$ \\ 
		\hline 
		$D_3$ & $15$ & $21$ & $0.101$ & $0.094$ & $0.094$ & $0.126$ & $0.113$ & $0.126$ & $0.113$ & $0.126$ & $0.107$ \\ 
		\hline 
		$D_4$ & $22$ & $28$ & $0.121$ & $0.096$ & $0.089$ & $0.127$ & $0.115$ & $0.127$ & $0.115$ & $0.102$ & $0.108$ \\ 
		\hline 
		$D_5$ & $29$ & $35$ & $0.133$ & $0.100$ & $0.093$ & $0.080$ & $0.120$ & $0.133$ & $0.120$ & $0.107$ & $0.113$ \\ 
		\hline 
		$D_6$ & $36$ & $41$ & $0.123$ & $0.097$ & $0.091$ & $0.130$ & $0.117$ & $0.110$ & $0.117$ & $0.104$ & $0.110$ \\ 
		\hline 
		$D_7$ & $42$ & $45$ & $0.116$ & $0.098$ & $0.091$ & $0.122$ & $0.110$ & $0.110$ & $0.122$ & $0.110$ & $0.122$ \\ 
		\hline \hline 
	\end{tabular} 
   }
   \caption{Structure of the dimension grid: expert weights regarding dimensions of the questionnaire along with the subdivision of items into dimensions.}
   \label{tab:dimBL}
\end{table}

\subsection{Applying 2TFLD method: first round}\label{section51}

To describe the application of the method in a simplified way, let us focus on one of the items that, in the first round, proved to be a point of conflict between the judges. This is item $I_{27}$ whose text in Spanish is found in the \emph{description} sheet: \emph{Considero que he alcanzado los objetivos del curso. Escala a utilizar: Tipo B}. In this case experts $J_1-J_3$ selected the linguistic term set $S^3$, expert $J_9$ set $S^5$ and experts $J_4 - J_8$ decided to perform the evaluations using $S^7$. The selection of the scales are given in the \emph{responses} sheet and in Table~\ref{tab:dataI27}. 

According to Section~\ref{section34} we have to undertake the following computational processes, which are:

\begin{itemize}
	\item \emph{Gathering phase}. 
	Table~\ref{tab:dataI27} shows the original assessments in consideration of ELH $h=3$, the matrix $(X_{27})_{9\times4}$. This information is stored in the \emph{responses} sheet. 
	
	\item \emph{2-tuple transformation phase}. 
	Linguistics values with respect to criterion $C_1$ to criterion $C_4$ are transformed into 2-tuples linguistic values by the application of Eq.~(\ref{eq:2Tuple}).
	
	\item \emph{Unification phase}. 
	A unified vision of data is achieved through the application of the transformation function given at Eq.~\ref{eq:transFunct}. In the particular case of $I_{27}$, Table~\ref{tab:unifiedI27} shows intermediate results $(X_{27}^{''})_{9\times4}$ of performing both transformational steps.
	
	\item \emph{Double aggregation phase}. We perform two rounds of aggregation with $\delta_1$ and $\delta_2$. According to expression Eq.~(\ref{eq:collective}) we obtain $(Y_{27})_{4}$, $W^{27}$ and $Z_{27}=(s_9, 0.263)$.
	
	\item \emph{Exploitation phase}. The main outcome of this phase is the re-translation of the linguistic solution to the 2-tuple-fuzzy-linguistic scale, $IS_{27}= (s_5, -0.369) \in S^7$. But also, the evaluation of the consensus degree obtained for this item. By using Eq.(\ref{eq:separation}) we have the following vector:
	$$
		\rho = \{7.680, 6.405, 4.481, 6.367, 5.861, 6.405, 1.994, 6.091, 9.182 \}
	$$
	Using $\rho$ we identify judges $J_1$ and $J_9$ as distant from the consensus. They may affect the consensus level  $CS_{27}=false$ because the consensus index obtained with Eq.(\ref{eq:CI}) is $CI_{27}=0.493$. With $\epsilon=0.75$ we get $RS_{27}=false$ and $RI_{27}=0.5$. 
	
\end{itemize}

Now the moderator analyzes with detail the data obtained with the use of \emph{2-tuple-fuzzy-delphi} tool in order to undertake modifications in $I_r$ and later by extension, into the full questionnaire. According to all the experts, the criteria are valuated as:
	
	$$
		Y_{27} = \{ (s_{11}, -0.121), (s_7, 0.255), (s_{11}, -0.070), (s_8, -0.013)\}
	$$

The item $I_{27}$ is quite well valued considering criterion $C_1$ and $C_3$, and the nine experts have considered that the relevance of this question in $Q$ is $W^{27}=0,987$. The value of $RI_{27}$ also tells moderator that is a good item, though not perfect. Thus $I_{27}$ is not rejected but modified. Considering that $C_2$ represents an evaluation about the writing, the text of this item is changed to \emph{Estoy satisfecho respecto al logro de los objetivos del curso. Escala a utilizar: Tipo B}. This new description is updated in the instance of the questionnaire in Google Forms, and also in the content of the \emph{description} sheet. 

\begin{table}[ht]
  \centering
	\begin{tabular}{c|c|c|c|c|c}
		\hline \hline
		\multicolumn{6}{c}{First round for $I_{27}$} \\ 
		\hline \hline
		Judge & Clarity & Writing & Presence & A.Scale & Relevance\\ 
		\hline 
		$J_1$ & $s_{2}^{3}$ & $s_{0}^{3}$ & $s_{2}^{3}$  & $s_{1}^{3}$ & $1.00$\\ 
		\hline 
		$J_2$ & $s_{2}^{3}$ & $s_{2}^{3}$ & $s_{2}^{3}$ & $s_{2}^{3}$ & $1.00$\\ 
		\hline 
		$J_3$ & $s_{2}^{3}$ & $s_{1}^{3}$ & $s_{2}^{3}$ & $s_{2}^{3}$ & $1.00$\\ 
		\hline 
		$J_4$ & $s_{5}^{7}$ & $s_{6}^{7}$ & $s_{6}^{7}$ & $s_{6}^{7}$ & $1.00$\\ 	
		\hline 
		$J_5$ & $s_{4}^{7}$ & $s_{3}^{7}$ & $s_{4}^{7}$ & $s_{2}^{7}$ & $0.90$\\ 
		\hline 
		$J_6$  & $s_{6}^{7}$ & $s_{6}^{7}$ & $s_{6}^{7}$ & $s_{6}^{7}$ & $1.00$\\ 
		\hline 
		$J_7$ & $s_{6}^{7}$ & $s_{3}^{7}$ & $s_{6}^{7}$ & $s_{4}^{7}$ & $1.00$\\ 
		\hline 
		$J_8$  & $s_{4}^{7}$ & $s_{4}^{7}$ & $s_{3}^{7}$ & $s_{3}^{7}$ & $1.00$\\ 
		\hline 
		$J_9$ & $s_{4}^{5}$ & $s_{1}^{5}$ & $s_{4}^{5}$ & $s_{0}^{5}$ & $0.99$\\ 
		\hline \hline 
	\end{tabular} 
  \caption{Gathered opinion regarding item $I_{27}$ considering the four linguistic criteria and the pertinence.}
  \label{tab:dataI27}
\end{table}

\begin{table}[ht]
	\centering
	\begin{tabular}{c|c|c|c|c}
		\hline \hline 
		Judge & Clarity & Writing & Presence & Scale \\ 
		\hline \hline
		$J_1$ &($s_{12}^{13}$,0) & ($s_{0}^{13}$,0) & ($s_{12}^{13}$,0) & ($s_{6}^{13}$,0) \\ 
		\hline 
		$J_2$ & ($s_{12}^{13}$,0) & ($s_{12}^{13}$,0) & ($s_{12}^{13}$,0) & ($s_{12}^{13}$,0)\\ 
		\hline 
		$J_3$ & ($s_{12}^{13}$,0) & ($s_{6}^{13}$,0) & ($s_{12}^{13}$,0) & ($s_{12}^{13}$,0) \\ 
		\hline 
		$J_4$ & ($s_{10}^{13}$,0) & ($s_{12}^{13}$,0) & ($s_{12}^{13}$,0) & ($s_{12}^{13}$,0) \\ 
		\hline 
		$J_5$ & ($s_{8}^{13}$,0) & ($s_{6}^{13}$,0) & ($s_{8}^{13}$,0) & ($s_{4}^{13}$,0)\\ 
		\hline 
		$J_6$ & ($s_{12}^{13}$,0) & ($s_{12}^{13}$,0) & ($s_{12}^{13}$,0) & ($s_{12}^{13}$,0) \\ 
		\hline 
		$J_7$ & ($s_{12}^{13}$,0) & ($s_{6}^{13}$,0) & ($s_{12}^{13}$,0) & ($s_{8}^{13}$,0) \\ 
		\hline 
		$J_8$ & ($s_{8}^{13}$,0) & ($s_{8}^{13}$,0) & ($s_{6}^{13}$,0) & ($s_{6}^{13}$,0) \\ 
		\hline 
		$J_9$ & ($s_{12}^{13}$,0) & ($s_{3}^{13}$,0) & ($s_{12}^{13}$,0) & ($s_{0}^{13}$,0) \\ 
		\hline \hline
	\end{tabular} 
	\caption{After the 2-tuple transformation and unification phases, the assessments are prepared for computing with words under the same scale $S^{13}$.}
	\label{tab:unifiedI27}
\end{table}

For the rest of the questionnaire, the expert panel gave several suggestions most of them addressed to grammar (\emph{use of plural and singular must match}), writing issues (were the case of $I_5, I_6, I_7, I_8$) and the answering scale (\emph{the "satisfied" scale doesn't match my positive impression}). Other comment frequently mentioned was: \emph{It is recommended that the wording of the question be homogeneous with respect to others}. This means that though the expert assess a single item each time, this person maintains an overall record of the questionnaire. Implies also that the last items of a dimension might be penalized in their valuations, not by the item itself (that may be perfectly formed and written), but because homogeneity. Thus the consistency in the style of writing could be considered as a new  criterion or as part of the instructions given to judges to consider in $C_2$.

In our case, the first round was most oriented to improve the wording, but still an early stage to detect consensus problems. For instance, everyone agreed that $I_{17}$ is not so reliable with $RI_{17}=0.25$. See the full description of this round in Table~\ref{tab:round1round2}.

\subsection{Applying 2TFLD method: second round}\label{section52}

The second iteration collects all the assessments given by the judges after receiving the new version $Q'$ and a document report with a description of the reliance and consensus status, $RS$ and $CS$ respectively, along with $IS_r$. To describe the second round of the 2TFLD method, we take up again the valuation of item $I_{27}$, and later we compare the output of the two rounds for the whole questionnaire.

To solve the MEMCLDM problem for $I_{27}$, the following computational processes are undertaken:

\begin{itemize}
	\item \emph{Gathering phase}. 
	Table~\ref{tab:dataI27round2} shows the original assessments $(X_{27})_{9\times4}$. This time everyone individually selected $S^7$.  
	\item \emph{2-tuple transformation phase}. 
	This step creates matrix $(X'_{27})_{9\times4}$ of 2-tuples linguistic values by the application of Eq.~(\ref{eq:2Tuple}).
	\item \emph{Unification phase}. 
	By the use of ELH aggregation operations happens in level $t^*=4$. After the application of $TF^{n(t)}_{t^*}$ (see Eq.~(\ref{eq:transFunct})), we get 
	$(X''_{27})_{9\times4}$, at it is given in Table~\ref{tab:unifiedI27round2}.
	\item \emph{Double aggregation phase}. Using operator $\delta_1$ we aggregate over the expert opinions, and using operator $\delta_2$ we aggregate over the criteria. We get $W^{27}=0.988$ and:
	$$
		Y _{27}~=~\{(s_{12}, 0), (s_{12}, -0.382) , (s_{12}, -0.255) , (s_{12},-0.217)\} 
	$$
	Now the item is best valorized with regards to the four linguistic criteria. 
	\item \emph{Exploitation phase}. 
	Here we re-translate $Z_{27}=(s_5, -0.396)$ to an upper level of the ELH as $TF{t^*}_{n(3)}(Z_{27})=(s_6, 0.104)=IS_{27}$. Again it is a better qualification, but we need to measure if everyone agrees with this result.  By using Eq.(\ref{eq:separation}) we have the following vector:
	$$
		\rho = \{0.254, 1.817, 0.254, 0.254, 0.254, 0.900, 0.254, 0.254, 0.921 \}
	$$
	Previous distant judges $J_1$ and $J_9$ now are close to the group. Only $J_2$ differs low. Using Eq.(\ref{eq:CI}), the consensus index is computed  
	as $CI_{27}=0.908$ (very close to $1$). Applying Eq.(\ref{eq:CI}) we get a reliance index of $RI_{27}=1$. So both markers are positive, $CS_{27}=true$ and $RS_{27}=true$. 
	
\end{itemize}

Related to the general performance of the questionnaire the previous situation is generalized: item scores are increased, consensus is achieved and reliance is validated. See the full description of this round in Table~\ref{tab:round1round2}. By comparison of round one and round two we can determine that $Q'$ is a consensual valid questionnaire for data collection regarding constructs: satisfaction in a community of inquiry and virtual communication in a community inquiry for a blended learning experience.

\begin{table}[ht]
	\centering
	\begin{tabular}{c|c|c|c|c|c}
		\hline \hline
		\multicolumn{6}{c}{Second round for $I_{27}$} \\ 
		\hline \hline
		Judge & Clarity & Writing & Presence & A.Scale & Pertinence\\ 
		\hline 
		$J_1$ & $s_{6}^{7}$ & $s_{6}^{7}$ & $s_{6}^{7}$  & $s_{6}^{7}$ & $1.00 $\\ 
		\hline 
		$J_2$ & $s_{6}^{7}$ & $s_{4}^{7}$ & $s_{6}^{7}$ & $s_{6}^{7}$ & $1.00 $\\ 
		\hline 
		$J_3$ & $s_{6}^{7}$ & $s_{6}^{7}$ & $s_{6}^{7}$ & $s_{6}^{7}$ & $1.00 $\\ 
		\hline 
		$J_4$ & $s_{6}^{7}$ & $s_{6}^{7}$ & $s_{6}^{7}$ & $s_{6}^{7}$ & $1.00 $\\ 
		\hline 
		$J_5$ & $s_{6}^{7}$ & $s_{6}^{7}$ & $s_{6}^{7}$ & $s_{6}^{7}$ & $0.99 $\\ 
		\hline 
		$J_6$  & $s_{6}^{7}$ & $s_{6}^{7}$ & $s_{5}^{7}$ & $s_{6}^{7}$ & $1.00 $\\ 
		\hline 
		$J_7$ & $s_{6}^{7}$ & $s_{6}^{7}$ & $s_{6}^{7}$ & $s_{6}^{7}$ & $1.00 $\\ 
		\hline 
		$J_8$  & $s_{6}^{7}$ & $s_{6}^{7}$ & $s_{6}^{7}$ & $s_{6}^{7}$ & $1.00 $\\ 
		\hline 
		$J_9$ & $s_{6}^{7}$ & $s_{6}^{7}$ & $s_{6}^{7}$ & $s_{5}^{7}$ & $0.90 $\\ 
		\hline \hline 
	\end{tabular} 
	\caption{Gathered opinion regarding item $I_{27}$ considering the four linguistic criteria and the numerical one.}
	\label{tab:dataI27round2}
\end{table}

\begin{table}[ht!]
  \centering
	\begin{tabular}{c|c|c|c|c}
		\hline \hline
		Judge & Clarity & Writing & Presence & A.Scale \\ 
		\hline 
		$J_1$ & $(s_{12}^{13}, 0)$ & $(s_{12}^{13}, 0)$ & $(s_{12}^{13}, 0)$  & $(s_{12}^{13}, 0)$ \\ 
		\hline 
		$J_2$ & $(s_{12}^{13}, 0)$ & $(s_{8}^{13}, 0)$ & $(s_{12}^{13}, 0)$ & $(s_{12}^{13}, 0)$ \\ 
		\hline 
		$J_3$ & $(s_{12}^{13}, 0)$ & $(s_{12}^{13}, 0)$ & $(s_{12}^{13}, 0)$ & $(s_{12}^{13}, 0)$\\ 
		\hline 
		$J_4$ & $(s_{12}^{13}, 0)$ & $(s_{12}^{13}, 0)$ & $(s_{12}^{13}, 0)$ & $(s_{12}^{13}, 0)$\\ 
		\hline 
		$J_5$ & $(s_{12}^{13}, 0)$ & $(s_{12}^{13}, 0)$ & $(s_{12}^{13}, 0)$ & $(s_{12}^{13}, 0)$\\ 
		\hline 
		$J_6$  & $(s_{12}^{13}, 0)$ & $(s_{12}^{13}, 0)$ & $(s_{10}^{13}, 0)$ & $(s_{12}^{13}, 0)$\\ 
		\hline 
		$J_7$ & $(s_{12}^{13}, 0)$ & $(s_{12}^{13}, 0)$ & $(s_{12}^{13}, 0)$ & $(s_{12}^{13}, 0)$\\ 
		\hline 
		$J_8$  & $(s_{12}^{13}, 0)$ & $(s_{12}^{13}, 0)$ & $(s_{12}^{13}, 0)$ & $(s_{12}^{13}, 0)$ \\ 
		\hline 
		$J_9$ & $(s_{12}^{13}, 0)$ & $(s_{12}^{13}, 0)$ & $(s_{12}^{13}, 0)$ & $(s_{10}^{13}, 0)$\\ 
		\hline \hline
	\end{tabular} 
  	\caption{Item $I_{27}$ unified assessments as 2-tuples linguistic values in round two.}
    	\label{tab:unifiedI27round2}
\end{table}

\begin{table}[H]
	\centering
	\resizebox{16cm}{!} {
		\begin{tabular}{c|ccccc|ccccc}
			\hline \hline
			& \multicolumn{5}{c}{$1^{er}$ Round} & \multicolumn{5}{c}{$2^{nd}$ Round} \\ 
			\hline \hline
			Item & $IS$ & $CS$ & $CI$ & $RS$ & $RI$ & $IS$ & $CS$ & $CI$ & $RS$ & $RI$\\ 
			\hline 
			$I_1$ & $(s_{5}^{7}, -0.182)$ & $true$ & $0.590$  & $false$ & $0.50$  & $(s_{6}^{7}, -0.370)$ & $true$ & $0.819$  & $true$ & $1.00$\\ 
			\hline 
			$I_2$ & $(s_{5}^{7}, -0,325)$ & $true$ & $0.544$  & $false$ & $0.50$  & $(s_{5}^{7}, 0.478)$ & $true$ & $0.758$  & $true$ & $1.00$\\ 
			\hline 
			$I_3$ & $(s_{4}^{7}, 0.0980)$ & $true$ & $0.528$  & $false$ & $0.25$  & $(s_{5}^{7}, 0.438)$ & $true$ & $0.728$  & $true$ & $1.00$\\ 
			\hline 
			$I_4$ & $(s_{5}^{7}, -0.311)$ & $true$ & $0.575$  & $true$ & $0.75$  & $(s_{6}^{7}, -0.360)$ & $true$ & $0.797$  & $true$ & $1.00$\\ 
			\hline 
			$I_5$ & $(s_{5}^{7}, -0.303)$ & $true$ & $0.563$  & $true$ & $1.00$  & $(s_{6}^{7}, -0.199)$ & $true$ & $0.863$  & $true$ & $1.00$\\ 
			\hline 
			$I_6$ & $(s_{5}^{7}, 0.056)$ & $true$ & $0.653$  & $true$ & $1.00$  & $(s_{6}^{7}, -0.272)$ & $true$ & $0.825$  & $true$ & $1.00$\\ 
			\hline 
			$I_7$ & $(s_{5}^{7}, -0.040)$ & $true$ & $0.637$  & $true$ & $1.00$  & $(s_{6}^{7}, -0.161)$ & $true$ & $0.883$  & $true$ & $1.00$\\ 
			\hline 
			$I_8$ & $(s_{5}^{7}, -0.020)$ & $true$ & $0.585$  & $true$ & $1.00$  & $(s_{6}^{7}, -0.219)$ & $true$ & $0.860$  & $true$ & $1.00$\\ 
			\hline 
			$I_9$ & $(s_{5}^{7}, 0.009)$ & $true$ & $0.543$  & $true$ & $0.75$  & $(s_{6}^{7}, -0.197)$ & $true$ & $0.871$  & $true$ & $1.00$\\ 
			\hline 
			$I_{10}$ & $(s_{4}^{7}, 0.347)$ & $false$ & $0.395$  & $true$ & $0.75$  & $(s_{6}^{7}, -0.341)$ & $true$ & $0.783$  & $true$ & $1.00$\\ 
			\hline 
			$I_{11}$ & $(s_{4}^{7}, 0.434)$ & $false$ & $0.438$  & $true$ & $0.75$  & $(s_{6}^{7}, -0.463)$ & $true$ & $0.772$  & $true$ & $1.00$\\ 
			\hline 
			$I_{12}$ & $(s_{5}^{7}, 0.031)$ & $true$ & $0.584$  & $true$ & $1.00$  & $(s_{6}^{7},-0.197)$ & $true$ & $0.864$  & $true$ & $1.00$\\ 
			\hline 
			$I_{13}$ & $(s_{5}^{7}, 0.188)$ & $true$ & $0.661$  & $true$ & $1.00$  & $(s_{6}^{7}, -0.372)$ & $true$ & $0.796$  & $true$ & $1.00$\\ 
			\hline 
			$I_{14}$ & $(s_{5}^{7}, -0.016)$ & $true$ & $0.575$  & $true$ & $0.75$  & $(s_{6}^{7}, -0.225)$ & $true$ & $0.850$  & $true$ & $1.00$\\ 
			\hline 
			$I_{15}$ & $(s_{5}^{7}, -0.382)$ & $true$ & $0.506$  & $false$ & $0.50$  & $(s_{6}^{7}, -0.385)$ & $true$ & $0.844$  & $true$ & $1.00$\\ 
			\hline 
			$I_{16}$ & $(s_{5}^{7}, -0.382)$ & $false$ & $0.466$  & $true$ & $1.00$  & $(s_{6}^{7}, -0.497)$ & $true$ & $0.789$  & $true$ & $1.00$\\ 
			\hline 
			$I_{17}$ & $(s_{4}^{7}, 0.483)$ & $true$ & $0.573$  & $false$ & $0.25$  & $(s_{5}^{7}, 0.404)$ & $true$ & $0.742$  & $true$ & $1.00$\\ 
			\hline 
			$I_{18}$ & $(s_{5}^{7}, -0.127)$ & $true$ & $0.563$  & $true$ & $1.00$  & $(s_{6}^{7}, -0.082)$ & $true$ & $0.933$  & $true$ & $1.00$\\ 
			\hline 
			$I_{19}$ & $(s_{5}^{7}, -0.399)$ & $true$ & $0.557$  & $false$ & $0.50$  & $(s_{6}^{7}, -0.388)$ & $true$ & $0.786$  & $true$ & $1.00$\\ 
			\hline 
			$I_{20}$ & $(s_{5}^{7}, 0.201)$ & $true$ & $0.651$  & $true$ & $0.75$  & $(s_{6}^{7}, -0.244)$ & $true$ & $0.832$  & $true$ & $1.00$\\ 
			\hline 
			$I_{21}$ & $(s_{5}^{7}, 0.261)$ & $true$ & $0.614$  & $true$ & $1.00$  & $(s_{6}^{7}, -0.132)$ & $true$ & $0.900$  & $true$  & $1.00$\\
			\hline
			$I_{22}$ & $(s_{5}^{7}, 0.295)$ & $true$ & $0.672$  & $true$ & $1.00$  & $(s_{6}^{7}, -0.115)$ & $true$ & $0.917$  & $true$  & $1.00$\\
			\hline
			$I_{23}$ & $(s_{4}^{7}, 0.358)$ & $false$ & $0.406$  & $false$ & $0.00$  & $(s_{5}^{7}, 0.468)$ & $true$ & $0.731$  & $true$  & $1.00$\\
			\hline
			$I_{24}$ & $(s_{5}^{7}, -0.013)$ & $true$ & $0.520$  & $true$ & $1.00$  & $(s_{6}^{7}, -0.328)$ & $true$ & $0.805$  & $true$  & $1.00$\\
			\hline
			$I_{25}$ & $(s_{5}^{7},-0.083)$ & $true$ & $0.536$  & $true$ & $1.00$  & $(s_{6}^{7}, -0.266)$ & $true$ & $0.818$  & $true$  & $1.00$\\
			\hline
			$I_{26}$ & $(s_{5}^{7},-0.068)$ & $true$ & $0.582$  & $true$ & $1.00$  & $(s_{6}^{7}, -0.287)$ & $true$ & $0.829$  & $true$  & $1.00$\\
			\hline
			$I_{27}$ & $(s_{5}^{7},-0.369)$ & $false$ & $0.493$  & $false$ & $0.50$  & $(s_{6}^{7}, -0.107)$ & $true$ & $0.908$  & $true$  & $1.00$\\
			\hline
			$I_{28}$ & $(s_{5}^{7},-0.430)$ & $false$ & $0.488$  & $true$ & $0.75$  & $(s_{6}^{7}, -0.306)$ & $true$ & $0.821$  & $true$  & $1.00$\\
			\hline
			$I_{29}$ & $(s_{4}^{7},0.320)$ & $false$ & $0.457$  & $false$ & $0.00$  & $(s_{6}^{7}, -0.225)$ & $true$ & $0.856$  & $true$  & $1.00$\\
			\hline
			$I_{30}$ & $(s_{5}^{7}, -0.073)$ & $true$ & $0.634$  & $true$ & $1.00$  & $(s_{6}^{7}, -0.263)$ & $true$ & $0.836$  & $true$  & $1.00$\\
			\hline
			$I_{31}$ & $(s_{5}^{7},-0.157)$ & $true$ & $0.584$  & $true$ & $0.75$  & $(s_{6}^{7}, -0.263)$ & $true$ & $0.836$  & $true$  & $1.00$\\
			\hline
			$I_{32}$ & $(s_{5}^{7},-0.235)$ & $true$ & $0.542$  & $true$ & $1.00$  & $(s_{6}^{7}, -0.202)$ & $true$ & $0.865$  & $true$  & $1.00$\\
			\hline
			$I_{33}$ & $(s_{5}^{7},-0.393)$ & $false$ & $0.466$  & $true$ & $1.00$  & $(s_{6}^{7}, -0.203)$ & $true$ & $0.873$  & $true$  & $1.00$\\
			\hline
			$I_{34}$ & $(s_{5}^{7}, -0.005)$ & $true$ & $0.630$  & $true$ & $0.75$  & $(s_{6}^{7}, -0.302)$ & $true$ & $0.829$  & $true$  & $1.00$\\
			\hline
			$I_{35}$ & $(s_{4}^{7}, 0.387)$ & $true$ & $0.532$  & $false$ & $0.50$  & $(s_{6}^{7}, -0.287)$ & $true$ & $0.818$  & $true$  & $1.00$\\
			\hline
			$I_{36}$ & $(s_{5}^{7}, 0.047)$ & $true$ & $0.624$  & $true$ & $1.00$  & $(s_{6}^{7}, -0.205)$ & $true$ & $0.868$  & $true$  & $1.00$\\
			\hline
			$I_{37}$ & $(s_{5}^{7}, 0.196)$ & $true$ & $0.612$  & $true$ & $1.00$  & $(s_{6}^{7}, -0.114)$ & $true$ & $0.917$  & $true$  & $1.00$\\
			\hline
			$I_{38}$ & $(s_{5}^{7}, -0.339)$ & $true$ & $0.538$  & $true$ & $0.75$  & $(s_{6}^{7}, -0.299)$ & $true$ & $0.813$  & $true$  & $1.00$\\
			\hline
			$I_{39}$ & $(s_{4}^{7},  0.352)$ & $true$ & $0.524$  & $false$ & $0.25$  & $(s_{6}^{7}, -0.323)$ & $true$ & $0.793$  & $true$  & $1.00$\\
			\hline
			$I_{40}$ & $(s_{5}^{7}, -0.075)$ & $true$ & $0.569$  & $true$ & $0.75$  & $(s_{6}^{7}, -0.117)$ & $true$ & $0.909$  & $true$ & $1.00$\\
			\hline
			$I_{41}$ & $(s_{4}^{7}, 0.442)$ & $true$ & $0.538$  & $false$ & $0.25$  & $(s_{6}^{7}, -0.291)$ & $true$ & $0.813$  & $true$ & $1.00$\\
			\hline
			$I_{42}$ & $(s_{4}^{7}, 0.203)$ & $false$ & $0.488$  & $false$ & $0.00$  & $(s_{6}^{7}, -0.279)$ & $true$ & $0.835$  & $true$ & $1.00$\\
			\hline
			$I_{43}$ & $(s_{5}^{7}, 0.326)$ & $true$ & $0.660$  & $true$ & $0.75$  & $(s_{6}^{7}, -0.102)$ & $true$ & $0.894$  & $true$ & $1.00$\\
			\hline
			$I_{44}$ & $(s_{5}^{7},-0.140)$ & $true$ & $0.581$  & $flase$ & $0.50$  & $(s_{6}^{7}, -0.271)$ & $true$ & $0.823$  & $true$ & $1.00$\\
			\hline
			$I_{45}$ & $(s_{5}^{7},-0.354)$ & $true$ & $0.507$  & $false$ & $0.50$  & $(s_{6}^{7}, -0.264)$ & $true$ & $0.803$  & $true$ & $1.00$\\
			\hline
			\hline \hline  
			\multirow{2}{*}{Q} & $CC$ & $CW$ & CP & $CAS$ & $CP$ & $CC$ & $CW$ & CP & $CAS$ & $CP$\\ 
			& $(s_{5}, -0.328)$ & $(s_{4}, 0.489)$ & $(s_{5}, -0.269)$ & $(s_{5}, -0.449)$ & $(s_{y}, y)$ & $(s_{6}, -0.426)$ & $(s_{5}, 0.497)$ & $(s_{6}, -0.449)$ & $(s_{6}, -0.390)$ &  $(s_{y}, y)$ \\ 
			\hline \hline 
		\end{tabular} 
	}
	\caption{Moderator compares first and second rounds. 2-tuples linguistic values are expressed under $S^7$.}
	\label{tab:round1round2}
\end{table}

\subsection{The 2-tuple-fuzzy-delphi DSS trimming tool}\label{section53}

Detailed data of the questionnaire linguistic score, the collective criteria assessments, and consensus or reliance status (such as those given in Table~\ref{tab:round1round2}) can be automatically acquired with the \emph{2-tuple-fuzzy-delphi} DSS, thus speeding up the time needed for understanding the results and making decisions. The moderator must undertake several decisions such as: (1) discarding an item with a low score and a high degree of consensus, (2) relocating items with a low pertinence value, or (3) re-writing the description of the item according to responses given to criteria. 

On certain occasions, it is recommend to reduce the number of questions, and this can be easily solved with this tool.
The \emph{2-tuple-fuzzy-delphi} DSS includes as a functionality a trim operator. In action, this operator filters by rows. The filtering criterion is to exceed a threshold label. Items with $IS_r$ below the threshold are hidden from the list. 

\begin{figure}[h!]
\begin{center}
\includegraphics[width=13.66cm,height=6.9cm]{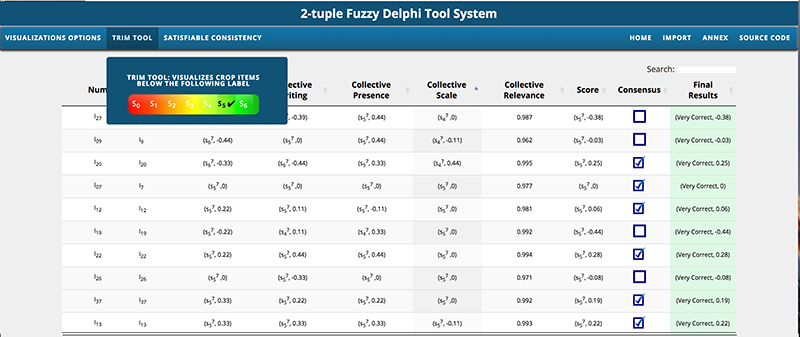}
\end{center}
\caption{The 2TFLD method is supported by an online tool. The trim operation helps the moderator to make decisions.}
\label{fig:trim}
\end{figure}

Let us suppose that our expert panel suggested for the second round a questionnaire $Q'$ up to $30$ items. We need to find the $15$ items worst rated in our questionnaire $Q$. This setting is shown in Figure~\ref{fig:trim}. We use a radio button input scale that considers $7$ linguistic terms and ranges from $s_0$ to $s_6$. This widget can be understood as seven buttons in one element. By clicking on button $s_1$, $s_2$ and subsequent, we are setting our threshold label. We found that selecting $s_5$ we can hide $17$ records and deliver $Q'$ with $n=28$. 

\section{Conclusions}\label{section6}

Innovative research hypotheses make hard to find standard data collection instruments already validated. Thus it is common to design a new questionnaire whose content validity can be tested through experts judgments by using the Delphi approach. The Fuzzy Delphi method is a recognized method for content validation. 

Our proposal, named 2-Tuple Fuzzy Linguistic Delphi (2TFLD) method, is an extension of the FD to use more information from the experts and to reflect over the expertise of the judge with the constructs covered. We allow the expert to give answers in the linguist scale that better suits its needs, by using extended linguistic hierarchies and the 2-tuple fuzzy linguistic model computation. A questionnaire of $n$ items has to solve $n$ instances of a Multi-Expert Multi-Criteria Linguistic Decision Making problem. Thus we compute individual item linguistic scores and collective linguistic information for the overall questionnaire using the deepest level of the hierarchy $S^{13}$, though they are simplified for better understanding to $S^7$ by or DSS tool. Consensus degrees and reliance levels are markers quite easy to understand by the moderator, along with the linguistic item score. All together offer valuable information about which items are of worst or better quality, after considering criteria such as clarity, writing, presence, answering scale or relevance.

We have conducted a blended learning experience in a high education scenario that combines FC and mL methodologies. To evaluate the experience of virtual communication between students-students and teacher-students using the Telegram app, a questionnaire was tested with a committee of nine experts. Mostly the panel of experts reflected a greater degree of knowledge or experience in one area or another, but not in combination. We acted as moderator and to validate the questionnaire we applied the 2TFLD method. The second iteration obtained consensus on all items. The final evaluation of the questionnaire was \emph{Excellent}. 

In this work there are several positive points that should be highlighted regarding the 2TFLD method:
\begin{itemize}
	\item It is innovative by resolving by CW processes, an otherwise too much simple process of binary acceptance / rejection. The panel of experts are high qualified researchers but their knowledge is almost discarded by a lousy expression. 
	\item It is generic, as the problem that it addressed is worldwide. 	
	\item It is flexible. The use of multigranular linguistic term sets, allows for comprehensible information given and information consumption. Also, it reflects the degree of expertise of each expert and for each dimension. Dimensions can match different research area in which not all the judges are experts at the same depth. 
	\item It is understandable. The use of linguistic values as the result for the valuation item by item, along with visual representations of the degree of reliance and consensus make it very easy to interpret if a consensus is reached to give content validity to the questionnaire.
	\item It is ready to use by the research community. We provide an online tool which is open source. Its main objective is to support the moderation in the task of testing a questionnaire. If the questionnaire is distributed to the expert panel using the Google Form solution, the responses can be almost directly imported in our tool.	
	\item It is helpful. We can apply it in our b-learning experience to listen the students. Other teachers can use this same questionnaire for the purpose of testing learning experiences that combine FC \& mL.
\end{itemize}

For future work, it is foreseen to extend our \emph{2tuple-fuzzy-delphi} DSS to integrate more functionality. This software may serve as a testbed for different multigranular term sets approaches, and for the proposal of new models for information fusion. Generally, we are interested in the exploitation and use of linguistic information in real-life problems to facilitate everyday decision making. We also consider to add user management to this tool in order to cover other areas of the Delphi method, such as the communication between the moderator and judges (for instance by sending the evaluation to the expert panel via e-mail), and to improve its usability.

\section{Acknowledgment}

This work was partly supported by the research project TIN2017-89517-P of the Ministry of Economy, cofinanciated by FEDER (European Regional Development Fund - ERDF).


\bibliographystyle{acm}      
\bibliography{teranga-go-preprint}   

\begin{thebibliography}{10}

\bibitem{AlEmran18}
{\sc Al-Emran, M., Mezhuyev, V., and Kamaludin, A.}
\newblock Technology acceptance model in m-learning context: A systematic
  review.
\newblock {\em Computers \& Education 125\/} (2018), 389--412.

\bibitem{Arbaugh08}
{\sc Arbaugh, J., Cleveland-Innes, M., Diaz, S., Garrison, D., Ice, P.,
  Richardson, J., and Swan, K.}
\newblock Developing a community of inquiry instrument: Testing a measure of
  the community of inquiry framework using a multi-institutional sample.
\newblock {\em The Internet and Higher Education 11}, 3-4 (2008), 133–136.

\bibitem{BARLETT1950}
{\sc Bartlett, M.~S.}
\newblock Tests of significance in factor analysis.
\newblock {\em British Journal of Mathematical and Statistical Psychology 3}, 2
  (1950), 77--85.

\bibitem{BERGMANN2012flip}
{\sc Bergmann, J., and Sams, A.}
\newblock {\em Flip your classroom: Reach every student in every class every
  day}.
\newblock International Society for Technology in Education, 2012.

\bibitem{berk1990importance}
{\sc Berk, R.~A.}
\newblock Importance of expert judgment in content-related validity evidence.
\newblock {\em Western journal of nursing research 12}, 5 (1990), 659--671.

\bibitem{CARRASCO2011linguistic}
{\sc Carrasco, R.~A., Villar, P., Hornos, M.~J., and Herrera-Viedma, E.}
\newblock A linguistic multi-criteria decision making model applied to the
  integration of education questionnaires.
\newblock {\em International Journal of Computational Intelligence Systems 4},
  5 (2011), 946--959.

\bibitem{CHANG2000}
{\sc Chang, P.-T., Huang, L.-C., and Lin, H.-J.}
\newblock The fuzzy delphi method via fuzzy statistics and membership function
  fitting and an application to the human resources.
\newblock {\em Fuzzy Sets and Systems 112}, 3 (2000), 511 -- 520.

\bibitem{chen1992fuzzy}
{\sc Chen, S.-J., and Hwang, C.-L.}
\newblock Fuzzy multiple attribute decision making methods.
\newblock In {\em Fuzzy multiple attribute decision making}. Springer, 1992,
  pp.~289--486.

\bibitem{CRONBACH1951}
{\sc Cronbach, L.~J.}
\newblock Coefficient alpha and the internal structure of tests.
\newblock {\em Psychometrika 16}, 3 (1951), 297--334.

\bibitem{dalkey1969}
{\sc Dalkey, N.}
\newblock An experimental study of group opinion: The delphi method.
\newblock {\em Futures 1}, 5 (1969), 408--426.

\bibitem{dalkey1963}
{\sc Dalkey, N., and Helmer, O.}
\newblock An experimental application of the delphi method to the use of
  experts.
\newblock {\em Management science 9}, 3 (1963), 458--467.

\bibitem{ding2002assessing}
{\sc Ding, C.~S., and Hershberger, S.~L.}
\newblock Assessing content validity and content equivalence using structural
  equation modeling.
\newblock {\em Structural Equation Modeling 9}, 2 (2002), 283--297.

\bibitem{Espinilla2011}
{\sc Espinilla, M., Liu, J., and Martinez, L.}
\newblock An extended hierarchical linguistic model for decision-making
  problems.
\newblock {\em Computational Intelligence 27}, 3 (2011), 489--512.

\bibitem{evers2013assessing}
{\sc Evers, A., Mu{\~n}iz, J., Hagemeister, C., H{\o}stm{\ae}lingen, A.,
  Lindley, P., Sj{\"o}berg, A., and Bartram, D.}
\newblock Assessing the quality of tests: Revision of the efpa review model.
\newblock {\em Psicothema 25}, 3 (2013), 283--291.

\bibitem{GARCIA201715}
{\sc García-Lira, K., Gutiérrez-Santiuste, E., and Montes-Soldado, R.}
\newblock Cuestionarios para la evaluacion de la comunicacion y la satisfaccion
  al aplicar metodologias flipped classroom combinadas con m-learning en
  educacion superior.
\newblock In {\em III Congreso Internacional de Educación Mediática y
  Competencia Digital\/} (2017), pp.~1145--1163.

\bibitem{GARRISON201384}
{\sc Garrison, D., and Akyol, Z.}
\newblock Toward the development of a metacognition construct for communities
  of inquiry.
\newblock {\em The Internet and Higher Education 17}, Supplement C (2013), 84
  -- 89.

\bibitem{GARRISON199987}
{\sc Garrison, D., Anderson, T., and Archer, W.}
\newblock Critical inquiry in a text-based environment: Computer conferencing
  in higher education.
\newblock {\em The Internet and Higher Education 2}, 2 (1999), 87 -- 105.

\bibitem{Herrera2009}
{\sc Herrera, F., Alonso, S., Chiclana, F., and Herrera-Viedma, E.}
\newblock Computing with words in decision making: Foundations, trends and
  prospects.
\newblock {\em Fuzzy Optimization and Decision Making 8}, 4 (Dec. 2009),
  337--364.

\bibitem{HERRERAYMARTINEZ2000}
{\sc Herrera, F., and Mart{\'\i}nez, L.}
\newblock A 2-tuple fuzzy linguistic representation model for computing with
  words.
\newblock {\em IEEE Transactions on fuzzy systems 8}, 6 (2000), 746--752.

\bibitem{HERRERAYMARTINEZ2001}
{\sc Herrera, F., and Mart{\'\i}nez, L.}
\newblock A model based on linguistic 2-tuples for dealing with multigranular
  hierarchical linguistic contexts in multi-expert decision-making.
\newblock {\em IEEE Transactions on fuzzy systems 31}, 2 (2001), 227–234.

\bibitem{hyrkas2003validating}
{\sc Hyrk{\"a}s, K., Appelqvist-Schmidlechner, K., and Oksa, L.}
\newblock Validating an instrument for clinical supervision using an expert
  panel.
\newblock {\em International Journal of nursing studies 40}, 6 (2003),
  619--625.

\bibitem{ISHIKAWA1993max}
{\sc Ishikawa, A., Amagasa, M., Shiga, T., Tomizawa, G., Tatsuta, R., and
  Mieno, H.}
\newblock The max-min delphi method and fuzzy delphi method via fuzzy
  integration.
\newblock {\em Fuzzy sets and systems 55}, 3 (1993), 241--253.

\bibitem{JALDEMARK2017editorial}
{\sc Jaldemark, J., and et~al.}
\newblock Editorial introduction: Collaborative learning enhanced by mobile
  technologies.
\newblock {\em British Journal of Educational Technology\/} (2017), 201--206.

\bibitem{KACPRZYK199221}
{\sc Kacprzyk, J., Fedrizzi, M., and Nurmi, H.}
\newblock Group decision making and consensus under fuzzy preferences and fuzzy
  majority.
\newblock {\em Fuzzy Sets and Systems 49}, 1 (1992), 21--31.

\bibitem{KAISER1974}
{\sc Kaiser, H.~F.}
\newblock An index of factorial simplicity.
\newblock {\em Psychometrika 39}, 1 (1974), 31--36.

\bibitem{Karabulut18}
{\sc Kaiser, H.~F.}
\newblock A systematic review of research on the flipped learning method in
  engineering education.
\newblock {\em British Journal of Educational Technology 49}, 3 (2018),
  398--411.

\bibitem{KIM2011763}
{\sc Kim, J.}
\newblock Developing an instrument to measure social presence in distance
  higher education.
\newblock {\em Br. J. of Educational Technol. 42}, 5 (2011), 763 -- 777.

\bibitem{LAGE2000inverting}
{\sc Lage, M.~J., Platt, G.~J., and Treglia, M.}
\newblock Inverting the classroom: A gateway to creating an inclusive learning
  environment.
\newblock {\em The Journal of Economic Education 31}, 1 (2000), 30--43.

\bibitem{LAWSHE1975}
{\sc Lawshe, C.~H.}
\newblock A quantitative approach to content validity.
\newblock {\em Personnel psychology 28}, 4 (1975), 563--575.

\bibitem{LISSITZ2007suggested}
{\sc Lissitz, R.~W., and Samuelsen, K.}
\newblock A suggested change in terminology and emphasis regarding validity and
  education.
\newblock {\em Educational researcher 36}, 8 (2007), 437--448.

\bibitem{LYNN1986}
{\sc Lynn, M.~R.}
\newblock Determination and quantification of content validity.
\newblock {\em Nursing research 35}, 6 (1986), 382--385.

\bibitem{MARTINEZ2012}
{\sc Martinez, L., and Herrera, F.}
\newblock An overview on the 2-tuple linguistic model for computing with words
  in decision making: Extensions, applications and challenges.
\newblock {\em Information Sciences 207\/} (2012), 1--18.

\bibitem{mendel2010computing}
{\sc Mendel, J.~M., Zadeh, L.~A., Trillas, E., Yager, R., Lawry, J., Hagras,
  H., and Guadarrama, S.}
\newblock What computing with words means to me [discussion forum].
\newblock {\em IEEE Computational Intelligence Magazine 5}, 1 (2010), 20--26.

\bibitem{Montes15}
{\sc Montes, R., Sanchez, A.~M., Villar, P., and Herrera, F.}
\newblock A web tool to support decision making in the housing market using
  hesitant fuzzy linguistic term sets.
\newblock {\em Applied Soft Computing 35\/} (2015), 949 -- 957.

\bibitem{Montes18}
{\sc Montes, R., Sanchez, A.~M., Villar, P., and Herrera, F.}
\newblock Teranga go!: Carpooling collaborative consumption community with
  multi-criteria hesitant fuzzy linguistic term set opinions to build
  confidence and trust.
\newblock {\em Applied Soft Computing 67\/} (2018), 941 -- 952.

\bibitem{morente18}
{\sc Morente-Molinera, J., Kou, G., Pérez, I., Samuylov, K., Selamat, A., and
  Herrera-Viedma, E.}
\newblock A group decision making support system for the web: How to work in
  environments with a high number of participants and alternatives.
\newblock {\em Applied Soft Computing 68\/} (2018), 191--201.

\bibitem{MURRAY1985}
{\sc Murray, T.~J., Pipino, L.~L., and van Gigch, J.~P.}
\newblock A pilot study of fuzzy set modification of delphi.
\newblock {\em Human Systems Management 5}, 1 (1985), 76--80.

\bibitem{NOORDERHAVEN1995strategic}
{\sc Noorderhaven, N.~G.}
\newblock {\em Strategic decision making}.
\newblock Wokingham: Addison-Wesley, 1995.

\bibitem{ORFANOU2015perceived}
{\sc Orfanou, K., Tselios, N., and Katsanos, C.}
\newblock Perceived usability evaluation of learning management systems:
  Empirical evaluation of the system usability scale.
\newblock {\em The International Review of Research in Open and Distributed
  Learning 16}, 2 (2015), 227--246.

\bibitem{PONTEROTTO2007}
{\sc Ponterotto, J.~G., and Ruckdeschel, D.~E.}
\newblock An overview of coefficient alpha and a reliability matrix for
  estimating adequacy of internal consistency coefficients with psychological
  research measures.
\newblock {\em Perceptual and Motor Skills 105}, 3 (2007), 997--1014.

\bibitem{RUBIN1999}
{\sc Rubin, S.~H.}
\newblock Computing with words.
\newblock {\em IEEE Transactions on Systems, Man, and Cybernetics, Part B
  (Cybernetics) 29}, 4 (1999), 518--524.

\bibitem{saint1994rules}
{\sc Saint, S., and Lawson, J.}
\newblock {\em Rules for reaching consensus: a modern approach to decision
  making}.
\newblock Amsterdam: Pfeiffer, 1994.

\bibitem{skjong2001expert}
{\sc Skjong, R., and Wentworth, B.~H.}
\newblock Expert judgment and risk perception.
\newblock In {\em The Eleventh International Offshore and Polar Engineering
  Conference\/} (2001), International Society of Offshore and Polar Engineers.

\bibitem{Maha76}
{\sc V.~Mahajan, H.A.~Linstone, M.~T.}
\newblock The delphi method: Techniques and applications.
\newblock {\em Journal of Marketing Research 13\/} (1976), 317.

\bibitem{WAN2013}
{\sc Wan, S.-P.}
\newblock 2-tuple linguistic hybrid arithmetic aggregation operators and
  application to multi-attribute group decision making.
\newblock {\em Knowledge-Based Systems 45\/} (2013), 31--40.

\bibitem{WEI2012}
{\sc Wei, G., and Zhao, X.}
\newblock Some dependent aggregation operators with 2-tuple linguistic
  information and their application to multiple attribute group decision
  making.
\newblock {\em Expert Systems with Applications 39}, 5 (2012), 5881--5886.

\bibitem{XU2011}
{\sc Xu, Y., and Wang, H.}
\newblock Approaches based on 2-tuple linguistic power aggregation operators
  for multiple attribute group decision making under linguistic environment.
\newblock {\em Applied Soft Computing 11}, 5 (2011), 3988--3997.

\bibitem{Yager2004}
{\sc Yager, R.~R.}
\newblock On the retranslation process in zadeh's paradigm of computing with
  words.
\newblock {\em IEEE Transactions on Systems, Man, and Cybernetics, Part B
  (Cybernetics) 34}, 2 (April 2004), 1184--1195.

\bibitem{ZADEH1965}
{\sc Zadeh, L.~A.}
\newblock Information and control.
\newblock {\em Fuzzy Sets 8}, 3 (1965), 338--353.

\bibitem{ZADEH1975}
{\sc Zadeh, L.~A.}
\newblock The concept of a linguistic variable and its application to
  approximate reasoning—i.
\newblock {\em Information Sciences 8}, 3 (1975), 199--249.

\bibitem{ZADEH1996}
{\sc Zadeh, L.~A.}
\newblock Fuzzy logic= computing with words.
\newblock {\em IEEE Transactions on Fuzzy Systems 4}, 2 (1996), 103--111.

\bibitem{zadeh2012computing}
{\sc Zadeh, L.~A.}
\newblock {\em Computing with words: Principal concepts and ideas}, vol.~277.
\newblock Springer, 2012.

\bibitem{Zhang2015}
{\sc Zhang, Y.~A.}
\newblock {\em Characteristics of Mobile Teaching and Learning}.
\newblock Springer Berlin Heidelberg, 2015.

\end{thebibliography}

\end{document}